\documentclass{aa}
\newif\ifreferee
\makeatletter
\@ifclasswith{aa}{referee}{\refereetrue}{\refereefalse}
\makeatother

\usepackage{txfonts}

\usepackage[T1]{fontenc}
\usepackage[english]{babel}
\usepackage[utf8]{inputenc}

\usepackage{amsmath,amsfonts,amssymb}
\usepackage{mathtools}
\usepackage{graphicx}
\usepackage{textcomp}
\usepackage{xcolor}
\usepackage[super]{nth}
\usepackage{mathtools}
\usepackage{multirow}
\usepackage{enumitem}
\usepackage{bigints}
\usepackage{verbatim}

\graphicspath{{figures/}} % Directory in which figures are stored
\usepackage[colorlinks=true, allcolors=blue]{hyperref}

\newcommand{\dd}{\mathrm{d}}
\newcommand{\dk}{\delta k}
\newcommand{\dkc}{\delta k_\mathrm{crit}}
\newcommand{\fc}{f_\mathrm{crit}}

\newcommand{\vect}[1]{\ensuremath{\mathbf{#1}}}

\usepackage{dblfloatfix}

\title{A correlation-locking adaptive filtering technique for minimum variance integral control in adaptive optics}
\ifreferee
	\titlerunning{CLOSE: a short title}
\fi

\begin{document}

\author{V.~Deo\inst{\ref{inst:LESIA},\ref{inst:Subaru}} \and%
	É.~Gendron\inst{\ref{inst:LESIA}} \and%
	F.~Vidal\inst{\ref{inst:LESIA}} \and%
	M.~Rozel\inst{\ref{inst:LESIA}} \and%
	A.~Sevin\inst{\ref{inst:LESIA}} \and%
	F.~Ferreira\inst{\ref{inst:LESIA}} \and%
	D.~Gratadour\inst{\ref{inst:LESIA},\ref{inst:ANU}} \and%
	N.~Galland\inst{\ref{inst:LESIA}} \and%
	G.~Rousset\inst{\ref{inst:LESIA}}
	}%
\institute{LESIA, Observatoire de Paris, Univ.~PSL, CNRS, Sorbonne Univ., Univ.~de Paris, 5 pl. Jules Janssen, 92195 Meudon, France\label{inst:LESIA}%
\and %
National Astronomical Observatory of Japan, Subaru Telescope, 650 North A'oh\=ok\=u Place, Hilo, HI 96720, U.S.A.\label{inst:Subaru} %
\and %
Research School of Astronomy and Astrophysics, Australian National University, Canberra, ACT 2611, Australia\label{inst:ANU} %
\\
email: \texttt{vdeo@naoj.org}
}

\date{Received 23 December 2020 / Accepted 17 March 2021}

\abstract{%
We propose the Correlation-Locking Optimization SchEme (CLOSE), a real-time adaptive filtering technique for adaptive optics (AO) systems controlled with integrators.
CLOSE leverages the temporal autocorrelation of modal signals in the controller telemetry and drives the gains of the integral command law in a closed servo-loop.
This supervisory loop is configured using only a few scalar parameters, and automatically controls the modal gains to closely match transfer functions achieving minimum variance control.
This optimization is proven to work throughout the range of noise and seeing conditions relevant to the AO system.

This technique has been designed while preparing the high-order AO systems for extremely large telescopes, in particular for tackling the optical gain (OG) phenomenon --a sensitivity reduction induced by on-sky residuals-- which is a prominent issue with pyramid wavefront sensors (PWFS).
CLOSE follows upon the linear modal compensation approach to OG, previously demonstrated to substantially improve AO correction with high order PWFS systems.
Operating on modal gains through multiplicative increments, CLOSE naturally compensates for the recurring issue of unaccounted sensitivity factors throughout the AO loop.

We present end-to-end simulations of the MICADO instrument single-conjugate AO to demonstrate the performances and capabilities of CLOSE.
We demonstrate that a single configuration shall provide an efficient and versatile optimization of the modal integrator while accounting for OG compensation, and while providing significant robustness to transient effects impacting the PWFS sensitivity.

%\red{Option 1:}
% LGS stuff - Finally, we demonstrate other stuff on LGS.
%\red{Option 2:}
%Finally, we show some day-time (on-sky) results obtained with the PWFS of the SCExAO high-contrast facility installed at Subaru Telescope, which (hopefully) confirm that CLOSE is good stuff.
}


\maketitle

%  ___       _             
% |_ _|_ __ | |_ _ __ ___  
% | || '_ \| __| '__/  _ \ 
% | || | | | |_| | | (_) |
% |___ |_| |_|\__|_| \___/ 

\begin{figure*}[b!]
	\centering
	\includegraphics[width=.75\textwidth]{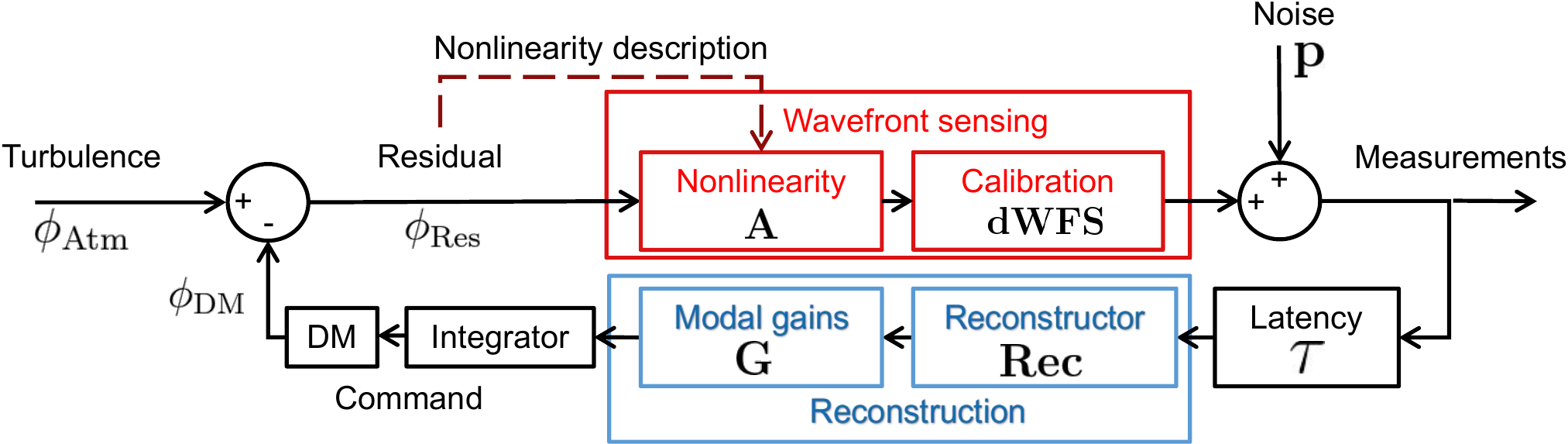}
	\caption{
		General modelization of the SCAO loop showing the WFS, real-time computer, and DM.
		The WFS modeled using the confusion matrix framework (Sect.~\ref{sec:2_1_og_mod}).
		The confusion matrix $\mathbf{A}$ is a random variable depending upon the residual wavefront $\phi_\mathrm{Res}$.
	}
	\label{fig:2_globalSchem}
\end{figure*}

\section{Introduction}
\label{sec:1_introduction}

Thanks to its extreme sensitivity as compared to other general-purpose wavefront sensors (WFSs) for astronomical adaptive optics (AO), the pyramid WFS (PWFS; \citealp{Ragazzoni1996Pupil})  has been the design choice included in all first light AO systems for the three upcoming extremely large telescopes (ELTs; \citealp{Tamai2018ELT, Fanson2018GMT, Liu2018TMT}), as well as a number of high performance systems on 8-10~m telescopes~\citep{Esposito2010First, Guyon2010SCExAO, Schatz2018Design}.

In recent years, the community has shown interest in tackling the non-linearity of the PWFS --the so-called \emph{optical gain} (OG)-- which is an on-sky sensitivity reduction induced by the limited dynamic range of the sensor, and can be modeled as a function of wavefront spatial frequency, with a magnitude depending on residual wavefront conditions, and therefore on ongoing turbulence statistics~\citep{Costa2005Modulation, Korkiakoski2008Applying, Deo2018Modal}.
When unmeasured and uncompensated, OG is a significant show-stopper that prevents the application of many algorithms relying upon the linearity of the servo-loop.
Two extremely common yet critical examples are: (1) the explicit estimation of the temporal transfer function, necessary for applying modal gain optimization techniques~\citep{Gendron1994Astronomical, Dessenne1998Optimization};
and (2) the proper subtraction of the non-common path aberration setpoint (tackled in, e.g., \citealp{Esposito2020NCPA}). For setpoint subtraction, the PWFS triggers a divergent positive feedback when attempting convergence to a wavefront setpoint which gradient exceeds the modulation radius.

Our previously proposed modal OG compensation pipeline~\citep{Deo2019Telescope} demonstrated significant performance improvements across different seeing conditions, but suffered from some limitations: it required introducing probe signals on the deformable mirror (DM); it also required a significant amount of preliminary modeling and computations; and it did not incorporate any modal variance minimization technique depending on noise level variations.
We propose in the present paper a novel algorithm, a self-regulating and entirely automated real-time modal gain controller, that resolves these three shortcomings: the Correlation-Locking Optimization SchemE (CLOSE).
The method is inspired from \cite{Montera2018Adaptive}, proposing to apply neural networks for --in particular-- tip-tilt sensitivity tracking and compensation. Preliminary results with CLOSE were published in~\citet{Deo2019CLOSE}.

CLOSE monitors the integrator overshoot through the temporal autocorrelation of modal WFS measurements, and drives modal gains through real-time multiplicative updates.
This allows for tracking and optimizing the sole temporal properties of the loop integrator; the OG compensation multipliers are automatically factored in the gain set into the command law, and do not require separate explicit computations.
The resulting steady-state command law may then be optimized regardless of the OG sensitivity reduction and can easily be tuned so as to compare to the minimum variance control of \citet{Gendron1994Astronomical}.

While we started designing CLOSE in the context of the single-conjugate AO (SCAO) module of the MICADO instrument~\citep{Davies2018MICADO, Clenet2019MICADO}, the technique naturally extends to a range of AO systems, namely for non-linear sensors performing well under a locally valid, modal, linearized description, such as the PWFS.
This includes other Fourier-like wavefront sensors~\citep{Fauvarque2019Kernel, Chambouleyron2020Convolution}, as well as  quad-cell Shack-Hartmanns, or the adaptation of a diffraction-limited, point source AO calibration into an adequate command law for extended natural sources or elongated laser stars, etc.

This paper is organized as follows; in Sect.~\ref{sec:2_model}, we present our formal model for analyzing the AO loop dynamics when using a WFS prone to optical gain.
In Sect.~\ref{sec:3_closeTheory}, we present the rationale for the CLOSE servo-loop, as well as semi-analytical demonstrations showing the achieved correspondence with minimum variance integrators.
Then we present in Sect.~\ref{sec:4_impl} two proposed implementations, real-time and offline, and briefly lay out the computational requirements associated with the use of CLOSE.
We show in Sect.~\ref{sec:5_results} the results of end-to-end numerical simulations, showing convergence, improved performance, and the breadth of applicable conditions for CLOSE on the MICADO SCAO system.
Finally, Sect.~\ref{sec:6_discu} includes some discussions on possible extensions and limitations of the scheme.

%  ______        _______ ____                        _      _ 
% |  _ \ \      / /  ___/ ___|   _ __ ___   ___   __| | ___| |
% | |_) \ \ /\ / /| |_  \___ \  | '_ ` _ \ / _ \ / _` |/ _ \ |
% |  __/ \ V  V / |  _|  ___) | | | | | | | (_) | (_| |  __/ |
% |_|     \_/\_/  |_|   |____/  |_| |_| |_|\___/ \__,_|\___|_|

\section{Modeling the AO loop}
\label{sec:2_model}
% Figure 1 has been move to main file because it wouldn't go where it belongs.

We show on Fig.~\ref{fig:2_globalSchem} the schematic of a SCAO loop that is considered throughout this paper, a model that has been designed to conveniently describe the OG effect on PWFSs operating in low Strehl regimes.
We will recall here the key elements relevant to this study; a more extensive description of this approach may be found in~\citet{Deo2019Telescope}.

The wavefronts --defined on Fig.~\ref{fig:2_globalSchem}-- $\phi_\mathrm{Atm}$, $\phi_\mathrm{DM}$ and $\phi_\mathrm{Res}$, while being continuous functions over the telescope aperture, are here implicitly meant as their decomposition over a control basis of the DM: $(\phi_1, ..., \phi_N)$ --where $N$ is the number of controlled modes--, plus an additional ``fitting'' component beyond the DM capabilities.
The left half of Fig.~\ref{fig:2_globalSchem} is represented in this modal space; the right half, where the measurement noise $\vect{p}$ is introduced, is in the WFS measurement space.
For a Shack-Hartmann, this would be the centroid displacement space, or for a PWFS, the space of either the normalized pixels or the gradient-like slopes maps.

The WFS is represented by the OG-describing matrix $\vect{A}$ (Sect.~\ref{sec:2_1_og_mod}), and the transformation from modal decomposition to measurements $\vect{dWFS}$: the modal interaction matrix of the wavefront sensor, defined as the Jacobian of the WFS response around the flat wavefront, i.e., computed using infinitesimal push-pulls.
Wavefront reconstruction is performed by matrix-vector multiplication, with the matrix $\vect{Rec}$ computed as the generalized inverse of $\vect{dWFS}$ --assuming the latter is adequately conditioned.
Finally, Fig.~\ref{fig:2_globalSchem} shows the temporal control of the loop is operated through a modal integrator, using a gain vector $\vect{G} = [G_i],_{1 \leq i \leq N}$.

\subsection{Optical gain: the confusion matrix model}
\label{sec:2_1_og_mod}

In the general case, the small-signal response of a nonlinear sensor is modified by the presence of atmospheric wavefront residuals.
The WFS Jacobian $\mathbf{dWFS}(\phi_\mathrm{Res})$ around any nonzero setpoint $\phi_\mathrm{Res}$ may be written as $\mathbf{dWFS}(\phi_\mathrm{Res}) = \mathbf{A}\cdot \mathbf{dWFS}$, as shown on Fig.~\ref{fig:2_globalSchem}.
In this description, the modal space operator $\vect{A}$ describes a local ``mixing'' of modal components around $\phi_\mathrm{Res}$ as compared to the calibrated response, and as such we call $\mathbf{A}$ the \emph{modal confusion matrix}.
The operator $\mathbf{A}$ is, generally, a random variable dependent on $\phi_\mathrm{Res}$.
In the simplest case, e.g. a Shack-Hartmann WFS with uniform centroid gain, the matrix $\vect{A}$ is a scaled identity matrix.

For the PWFS case, it has been shown that the confusion matrix has some reasonable properties when described on an appropriate modal basis, which are the foundation of optical gain modal compensation for the PWFS \citep{Korkiakoski2008Improving}.
In previous work \citep{Deo2018Modal, Deo2019Telescope}, we performed a thorough numerical assessment of the fluctuations of $\vect{A}$ when the spatial power spectrum density (PSD) of $\phi_\mathrm{Res}$ is stationary.
These analyses were performed using a Karhunen-Loève (KL) basis orthonormalized on the DM \citep{Ferreira2018Adaptive}.
This basis is made of modes containing an isotropic mix of spatial frequencies of a single norm, sorted by increasing frequency. The last \textasciitilde300 modes (out of $4\,300$ total modes for the ELT), which structure is impacted by the DM cutoff, contain a variety of waffles.
We use this basis for all purposes in this paper.

When using a decomposition on our DM KL basis, we have previously demonstrated that: (1) $\vect{A}$ is essentially diagonal for low-order modes, which bear most of the power of the atmospheric turbulence; (2) that its diagonal coefficients vary by no more than a few percent for a given set of wavefronts $\phi_\mathrm{Res}$ of identical PSD, a property in accordance with convolutional PWFS descriptions \citep{Fauvarque2019Kernel, Chambouleyron2020Convolution}; and (3) that the off-diagonal portion of $\vect{A}$, while non-negligible for high-order $\phi_i$, is of negligible statistical average.

Using such a basis which statistically diagonalizes $\vect{A}$ enables the modal gain compensation strategy, as shown in the ``Reconstruction'' block of Fig.~\ref{fig:2_globalSchem}.
The modal gains set in $\vect{G}$ cover two functionalities: first, they include the compensation of $\vect{A}$ for ongoing turbulent conditions; and second they define the transfer function gain such that the modal integrators exhibit appropriate rejection levels.
Both of these terms are ambiguously mixed into the coefficients of $\vect{G}$ set into the controller.
This ambiguity will impact the dynamical modeling of the system, laid out in Sect.~\ref{sec:2_3_transferFunctions}, as well as the optimization strategies explained in Sect.~\ref{sec:3_closeTheory}.
Its relationship to proper non common path aberration compensation, which requires independently identifying the compensation of $\vect{A}$, is discussed in Sect.~\ref{sec:6_2}.

The stability of $\vect{A}$ against the PSD of $\phi_\mathrm{Res}$ ensures that adequate $G_i$ values vary with the same temporal scale than the descriptive statistical parameters ($r_0$, $L_0$, $C_n^2(h)$, ...) of the turbulence.

\subsection{Diagonalizing the loop model}
\label{sec:2_2_diagonalizing}

\begin{figure}[t!]
	\centering
	\ifreferee
		\includegraphics[width=.59\textwidth]{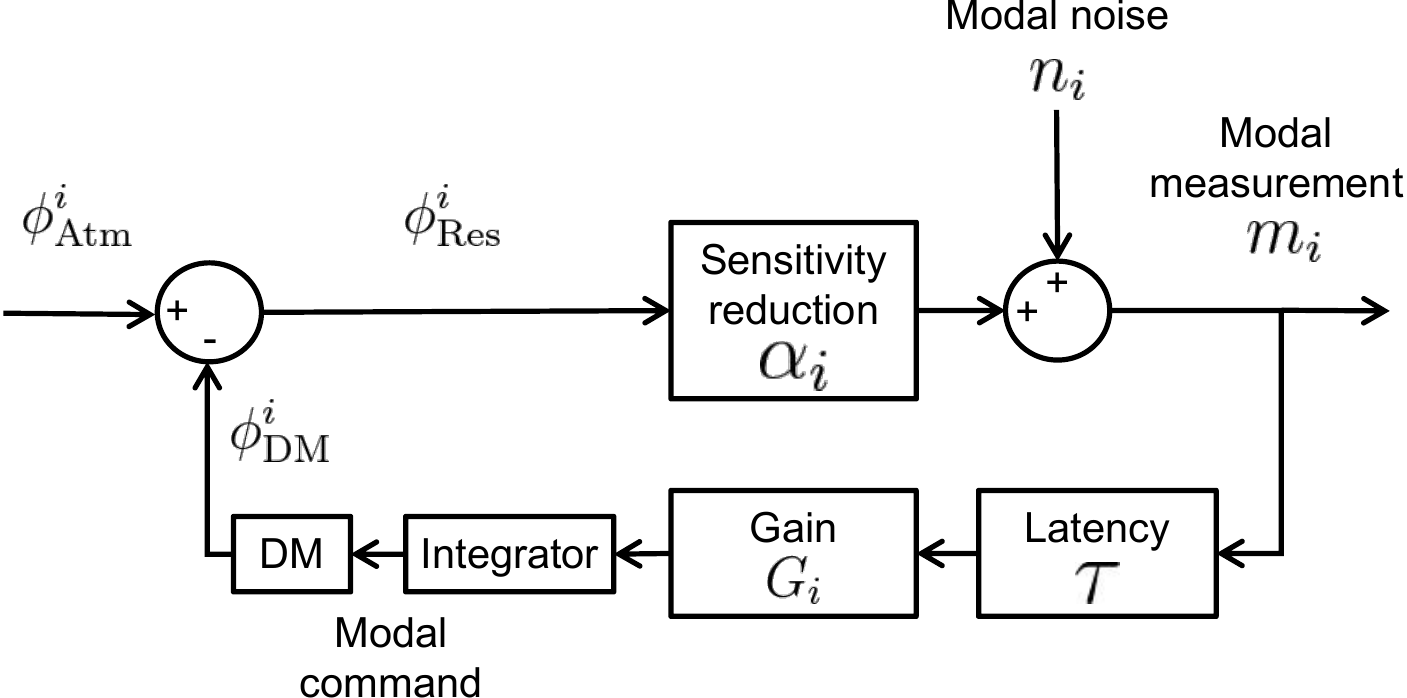}%
	\else
		\includegraphics[width=.99\columnwidth]{2_modalschem.pdf}%
	\fi
	\makeatother
	\caption{%
		Diagonalization of the AO loop model (Fig.~\ref{fig:2_globalSchem}) for mode $i$, under the approximation that $\vect{A}$ is diagonal, and thus reduced to its $i-$th diagonal term $\alpha_i$.
	}
	\label{fig:2_modalSchem}
\end{figure}

Following through with the approximation of statistical diagonality of $\mathbf{A}$, Fig.~\ref{fig:2_globalSchem} can then be simplified to the flowchart shown on Fig.~\ref{fig:2_modalSchem}, as the reference interaction and reconstruction matrices $\vect{dWFS}$ and $\vect{Rec}$ reduce to an identity matrix.
Fig.~\ref{fig:2_modalSchem} applies as one of $N$ decoupled servo-loops for each of the controlled modes, with $\vect{A}$ reduced to its $i-$th diagonal coefficient $\alpha_i$. We call $\alpha_i$ the modal sensitivity \emph{reduction} coefficient, as $0 < \alpha_i \leq 1$ stands for a PWFS.
With $\alpha_i$ varying only with the statistical properties of the turbulence, it can be considered as a constant for the purposes of analyzing the model in Fig.~\ref{fig:2_modalSchem} for stationary or slowly evolving atmospheric conditions.
With those hypotheses, Fig.~\ref{fig:2_modalSchem} is no more than a most classical integral feedback loop, but for the unknown and unmeasured $\alpha_i$.

Furthermore, our choice of basis is descriptively similar to atmospheric KL modes~\citep{Dai1996Modal} for the modes that bear the most of the turbulence power.
%With our KL basis being close to diagonalizing the covariance of the input $\phi_\mathrm{Atm}$ (and the measurement noise being white)
Thus, the $N$ modal loops as shown in Fig.~\ref{fig:2_modalSchem} may be approximated as uncorrelated, with their independent optimizations resulting in a global one.

\subsection{Transfer functions}
\label{sec:2_3_transferFunctions}

Each of the $N$ modal servo-loops is entirely described by a small number of parameters: the temporal spectra of the turbulence and noise for the $i$-th mode, and the $\alpha_i$ (unknown) and $G_i$ (known) scalars.
The atmospheric temporal spectrum for a single KL mode is well described in the literature \citep{Conan1995Wavefront, Gendron1995Optimisation} for Kolmogorov or Von Kármán turbulence:
\begin{equation}
\label{eq:2_3_turbSpectrum}
|\hat{\phi}_\mathrm{Atm}^i(f)|^2 \propto\! f^0 \text{ for } f < f_i\text{, and}%
\propto\! f^{-17/3} \text{ for } f > f_i,
\end{equation}
where $f$ denotes temporal frequencies, $\hat{\bullet}$ temporal Fourier transforms, and $f_i$ is the mode cutoff frequency, given by the wind speed and the mode radial order.
We will use the spectrum of Eq.~\ref{eq:2_3_turbSpectrum} as a reference template for the analyses led in Sect.~\ref{sec:3_closeTheory}.

The minimization objective we consider for integral control optimization is the variance of the residual:
\begin{equation}
\label{eq:2_3_PSDVariance}
\mathrm{Var}_k \left(\phi_\mathrm{Res}^i[k] \right) = \int_f \left|\hat{\phi}_\mathrm{Res}^i(f)\right|^2\dd f,
\end{equation}
where $k$ is the temporal sample index.
Yet, one cannot access directly the value of the $\phi_\mathrm{Res}^i[k]$, but only those of the measurements $m_i[k]$. To perform semi-analytical computations of how to achieve this minimization, as we will do in Sect.~\ref{sec:3_3_CLOSE_MV_comparisons}, it is necessary to introduce the various transfer functions between the quantities involved.

Those can be easily obtained (e.g.,~\citealp{Madec1999Wavefront}) from Fig.~\ref{fig:2_modalSchem}, and we here recall their expressions in a time-sampled framework.
A discussion on continuous vs. discrete time approaches to AO is proposed in Appendix~\ref{appdx:C_CLOSE_TransferFunc}; it needs only be noted here that discrete-time is valid \citep{Kulcsar2006Optimal} as (1) temporal aliasing is negligible, which is the case from the rapid decrease of Eq.~\ref{eq:2_3_turbSpectrum}; and (2) frequencies involved are well below Nyquist.
Let us define the shorthand (Eq.~\ref{eq:C_hcorrdts}), where $T$ is the sampling period and $j^2$=-1:
\begin{equation}
\label{eq:2_hi}
h(f;g) = \left[1 + g\cdot \dfrac{\exp(-2j \pi f (T+\tau))}{1-\exp(-2j \pi f T)} \right]^{-1},
\end{equation}
using which we express the transfer functions, where the subscripts identify the input (either $\phi^i_\mathrm{Atm}$ or the modal noise $n_i$) and output ($\phi^i_\mathrm{Res}$ or the modal measurement $m_i$) considered:
\begin{align}
\label{eq:2_HatmToRes}
h^i_{\mathrm{Atm}\longrightarrow\mathrm{Res}} & = h(f; \alpha_i G_i),                                                     \\
\label{eq:2_HnToRes}
h^i_{\mathrm{n}\longrightarrow\mathrm{Res}}   & = - \dfrac{1}{\alpha_i}\dfrac{\exp(-2j \pi f (T+\tau))}{1-\exp(-2j \pi f T)}~h(f;\alpha_i G_i),
\end{align}
and
\begin{align}
\label{eq:2_HatmToM}
h^i_{\mathrm{Atm}\longrightarrow\mathrm{m}} & = \alpha_i\  h(f;\alpha_i G_i) \\
\label{eq:2_HnToM}
h^i_{\mathrm{n}\longrightarrow\mathrm{m}}   & = h(f;\alpha_i G_i).
\end{align}
These transfer functions are formally similar to those of an AO loop unaffected by optical gain ($\alpha_i = 1$), except for an effective noise amplification, as seen from the $-1/\alpha_i$ in Eq.~\ref{eq:2_HnToRes}.
It follows from Eqs.~\ref{eq:2_HatmToM} and~\ref{eq:2_HnToM} that the Fourier transform of the measurements can be written as:
\begin{equation}
\label{eq:2_eqtransfer}
\hat{m}_i(f) = h(f; \alpha_iG_i)\left(\alpha_i\ \hat{\phi}_\mathrm{Atm}^i(f) + \hat{n}_i(f)\right).
\end{equation}
Eq.~\ref{eq:2_eqtransfer} emphasizes that for a given turbulence PSD, $\hat{m}_i(f)$ is entirely defined by two parameters:
the transfer function gain $\alpha_i G_i$, and the sensitivity-adjusted signal-to-noise ratio (S/N), noted $\sigma_i$:
\begin{equation}
\label{eq:2_sigma_i}
\sigma_i^2 = {\alpha_i}^2\ \dfrac{\mathrm{Var} \left(\phi_\mathrm{Atm}^i \right)}{\mathrm{Var} \left(n_i \right)}.
\end{equation}

Minimum variance control involves setting $G_i$ such that $\alpha_i G_i$ is the minimizer of the quantity expressed in Eq.~\ref{eq:2_3_PSDVariance} for the given adjusted S/N.
This either requires measuring $\alpha_i$ explicitly, as is done for instance in other optical gain compensation methods \citep{Korkiakoski2008Improving, Esposito2015Non, Esposito2020NCPA}, or finding a proxy to indirectly infer the hidden value $\alpha_i G_i$, and adjust $G_i$ adequately.

%   ____ _     ___  ____  _____                   _   __  ____     __
%  / ___| |   / _ \/ ___|| ____|   __ _ _ __   __| | |  \/  \ \   / /
% | |   | |  | | | \___ \|  _|    / _` | '_ \ / _` | | |\/| |\ \ / / 
% | |___| |__| |_| |___) | |___  | (_| | | | | (_| | | |  | | \ V /  
%  \____|_____\___/|____/|_____|  \__,_|_| |_|\__,_| |_|  |_|  \_/   

%\clearpage
\section{The correlation-locking scheme}
\label{sec:3_closeTheory}
%The remaining parameter to entirely describe the loop is the effective signal-to-noise ratio (SNR): the turbulence to noise power ratio, amplified by $\alpha_i^{-1}$.

When the modal sensitivity reductions $\alpha_i$ are unknown, the closed-loop measurements $m_i$ do still contain sufficient information to retrieve $h(f; \alpha_iG_i)$ and control $G_i$ towards optimal rejection.
We intend to perform this control without computing an extensive spectrum of the measurements, nor performing an explicit estimation of the system response, but leveraging the short-term temporal autocorrelation of the measurements as a proxy for  the effective modal loop gain $\alpha_i G_i$.
We present in this section the Correlation-Locking optimization SchemE (CLOSE), which drives the modal gains $G_i$ in real-time towards a favorable solution for rejection.
This section covers the foundation and steady-state solutions for the loop transfer functions achieved using CLOSE; we will then present in Sect.~\ref{sec:4_impl} how we implement CLOSE as a second-layer supervisory loop, taking as inputs the modal WFS measurements and operating on the gain vector.

\subsection{Rationale: Using the loop resonance}
\label{sec:3_1_rationale}

\begin{figure}[t]
	\centering
	\ifreferee
	\includegraphics[width=0.65\columnwidth]{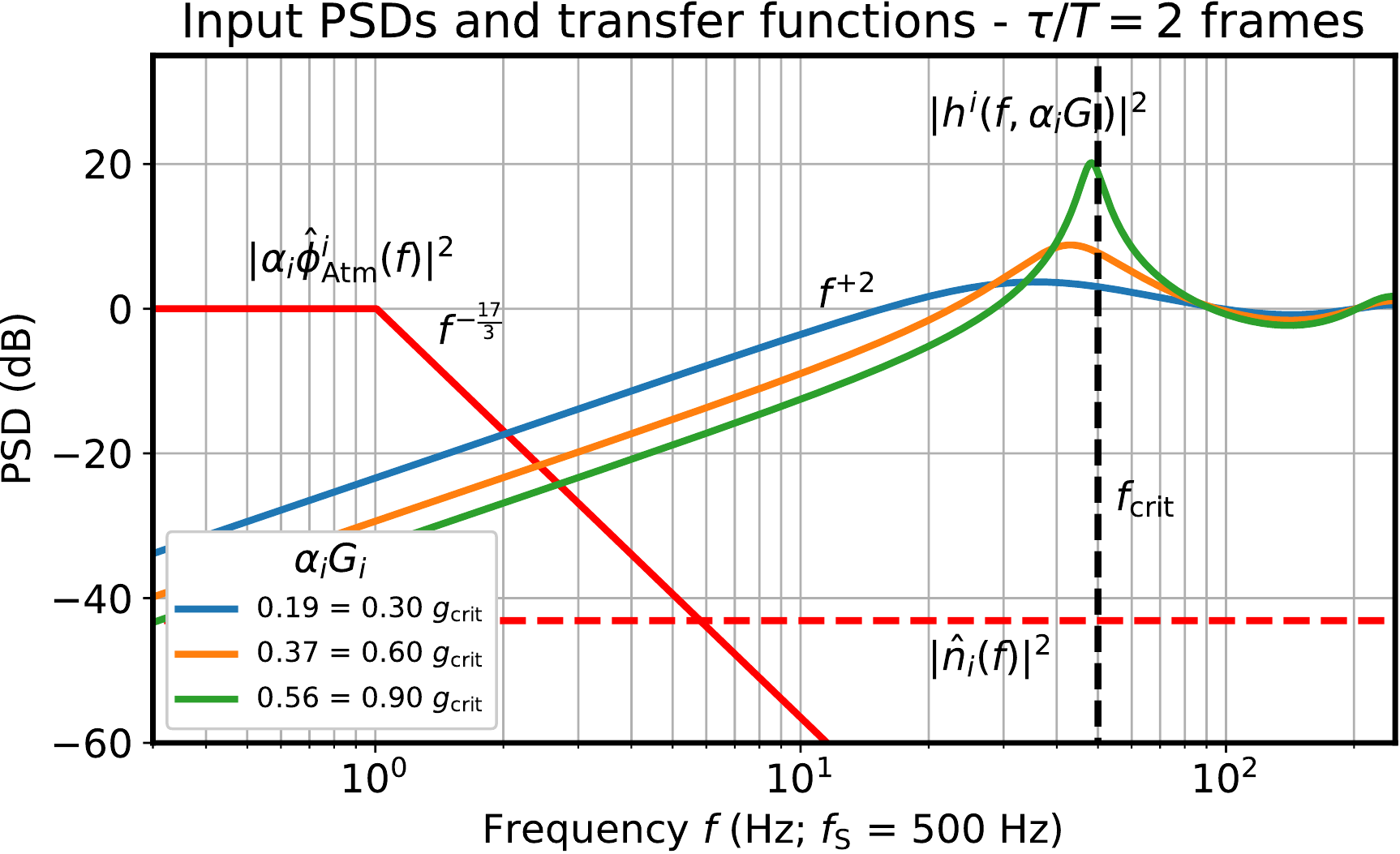}\\[.5em]
	\includegraphics[width=0.65\columnwidth]{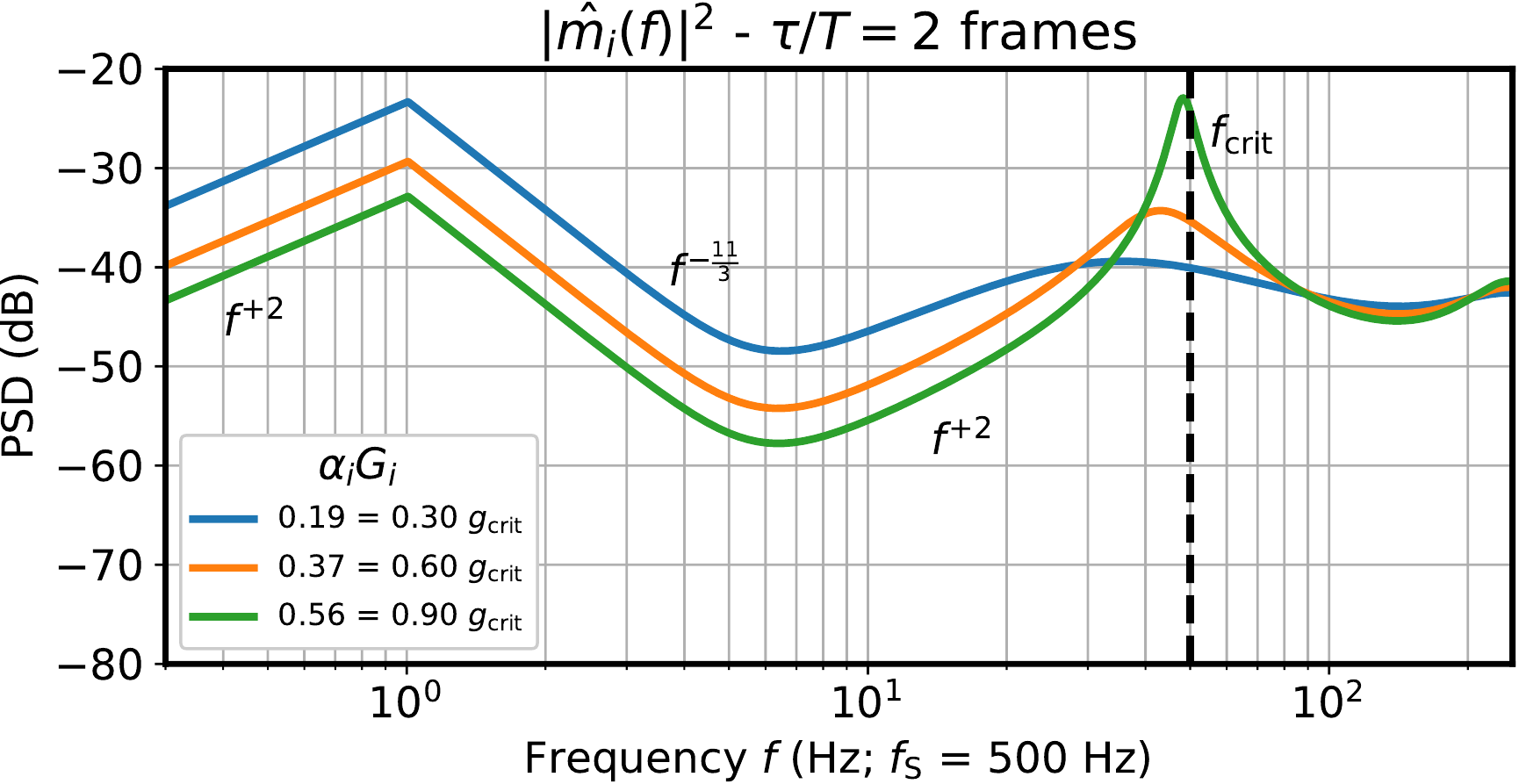}
	\else
	\includegraphics[width=0.98\columnwidth]{3_transferFuncs1.pdf}\\[.5em]
	\includegraphics[width=0.98\columnwidth]{3_transferFuncs2.pdf}
	\fi
	\caption{Typical spectral components for our AO semi-analytical computations.
		Top: input spectra for turbulence and noise, and the measurement power transfer function $|h^i(f;\alpha_i G_i)|^2$ for three $\alpha_i G_i$ values.
		Bottom: corresponding measurement power spectra $|\hat{m}_i(f)|^2$.
		For this example, $\tau = 2T$, and $f_S = 500$~Hz, yielding $g_\mathrm{crit} \approx 0.61$ and $f_\mathrm{crit}=50$~Hz.
	}
	\label{fig:3_spectralOutput}
\end{figure}

For each mode, the loop described on Fig.~\ref{fig:2_modalSchem} is a classical feedback loop with an integral controller of gain $\alpha_i G_i$ and total delay\footnote{%
% FOOTNOTE
\emph{Expected value} of the delay between the occurrence of a perturbation $\phi_\mathrm{Atm}(t)$ and the mean time of its correction on the DM.
% END FOOTNOTE
} $\tau + T$.
The transfer function $h(f; g)$, as developed in Sect.~\ref{sec:2_3_transferFunctions}, is a well characterized high-pass filter, with a $f^{+2}$ square modulus up to the roll-off, followed by a resonance peak located beyond.
Some examples of $h(f; g)$, using $\tau = 2T$, are shown on Fig.~\ref{fig:3_spectralOutput} (top).
$h(f;g)$ is a stable filter up to a maximum gain value $g=g_\mathrm{crit}$, which depends only on the normalized latency $\dfrac{\tau}{T}$.
As the gain $g$ increases towards $g_\mathrm{crit}$, the resonant peak sharpens and its amplitude increases; the peak central frequency increases and converges towards the critical frequency $f_\mathrm{crit}$.
In other terms, the Laplace transform associated to $h(f;g)$ sees its lowest-frequency pole displacing towards the imaginary axis, eventually intercepting it at $s = 2j\pi f_\mathrm{crit}$ for $g = g_\mathrm{crit}$.
Values of $f_{crit}$ and $g_\mathrm{crit}$ with latency are provided in Table~\ref{tab:3_summary} for reference; the given general formula is demonstrated in Appendix~\ref{appdx:C_CLOSE_TransferFunc}.

\begin{table}[h!]
	\centering
	\caption{%
		Parameters related to transfer function divergence depending on the latency of the system. 	
		$f_S$ is the sampling rate; $\dkc = f_S\,/\,2\fc$ is discussed in Sect.~\ref{sec:3_2_CLOSE}.
	}
	\label{tab:3_summary}
	\begin{tabular}{cccc}
		\hline
		\hline
		$\tau\,/\,T$ & $f_\mathrm{crit}\,/\,f_S$ & $g_\mathrm{crit}$ & $\dkc$ \\
		(frames) & & & (frames) \\
		\hline
		0 & $1~/~2$ & 2.0 & 1\\
		1 & $1~/~6$ & 1.0 & 3\\
		2 & $1~/~10$ & $\approx0.618$ & 5\\
		$\dfrac{\tau}{T}$ & $1~/~\left(4\dfrac{\tau}{T}+2\right)$ & $2\sin \left( \pi \dfrac{\fc}{f_S}\right) $ & $2\dfrac{\tau}{T}+1$\\
		\hline
	\end{tabular}
\end{table}

The behavior of the rejection peak as the transfer function approaches divergence is illustrated on Fig.~\ref{fig:3_spectralOutput}, showing the PSDs for the atmospheric mode and the noise, and some examples of transfer functions and output spectra $|\hat{m}_i(f)|^2$ for three values of $g=\alpha_i G_i$.
The latency is $\tau=2T$, and the sensitivity adjusted S/N (Eq.~\ref{eq:2_sigma_i}) was taken as $\sigma_i=10$.
Without loss of generality, we normalized the atmospheric spectrum at $|\alpha_i\hat{\phi}^i_\mathrm{Atm}(f=0)|^2 = 1$.

The atmospheric spectrum, filtered by the high-pass transfer function, results in the leftmost peak of $|\hat{m}_i(f)|^2$ seen on Fig.~\ref{fig:3_spectralOutput} (bottom), at the modal cutoff frequency $f_i=1$~Hz.
Near 10~Hz, we see a transition from rejected turbulence, showing a $f^{-11/3}$ spectrum, into a regime of $f^{+2}$, as the input $\alpha_i \phi^i_\mathrm{Atm} + n_i$ is dominated by the measurement noise.
This regime is followed by the rejection peak, which amplifies and shifts towards $f_\mathrm{crit} = 50$~Hz as the gain increases.

\subsection{Correlation locking: steady state objective}
\label{sec:3_2_CLOSE}

We assume that the AO latency $\tau$ has been calibrated and is a fixed, known parameter; consequently, so is $\fc$.
For reasonable $\tau$ values, $f_\mathrm{crit}$ lies in the noise floor of the spectrum, well beyond the turbulence cutoff frequency.
In its simplest form, our philosophy is to notice that the amplitude of the resonant peak may be used as a proxy for the effective modal gain $\alpha_i G_i$.

We have however deemed that it is particularly inconvenient to attempt real-time estimations of the peak amplitude or structure --which calls for extensive buffer acquisitions, explicit PSD estimations, etc.
The latter remains well within technical reach, and may be done at later stages of this research.

\begin{figure}
	\centering
	\ifreferee
		\includegraphics[width=.65\columnwidth]{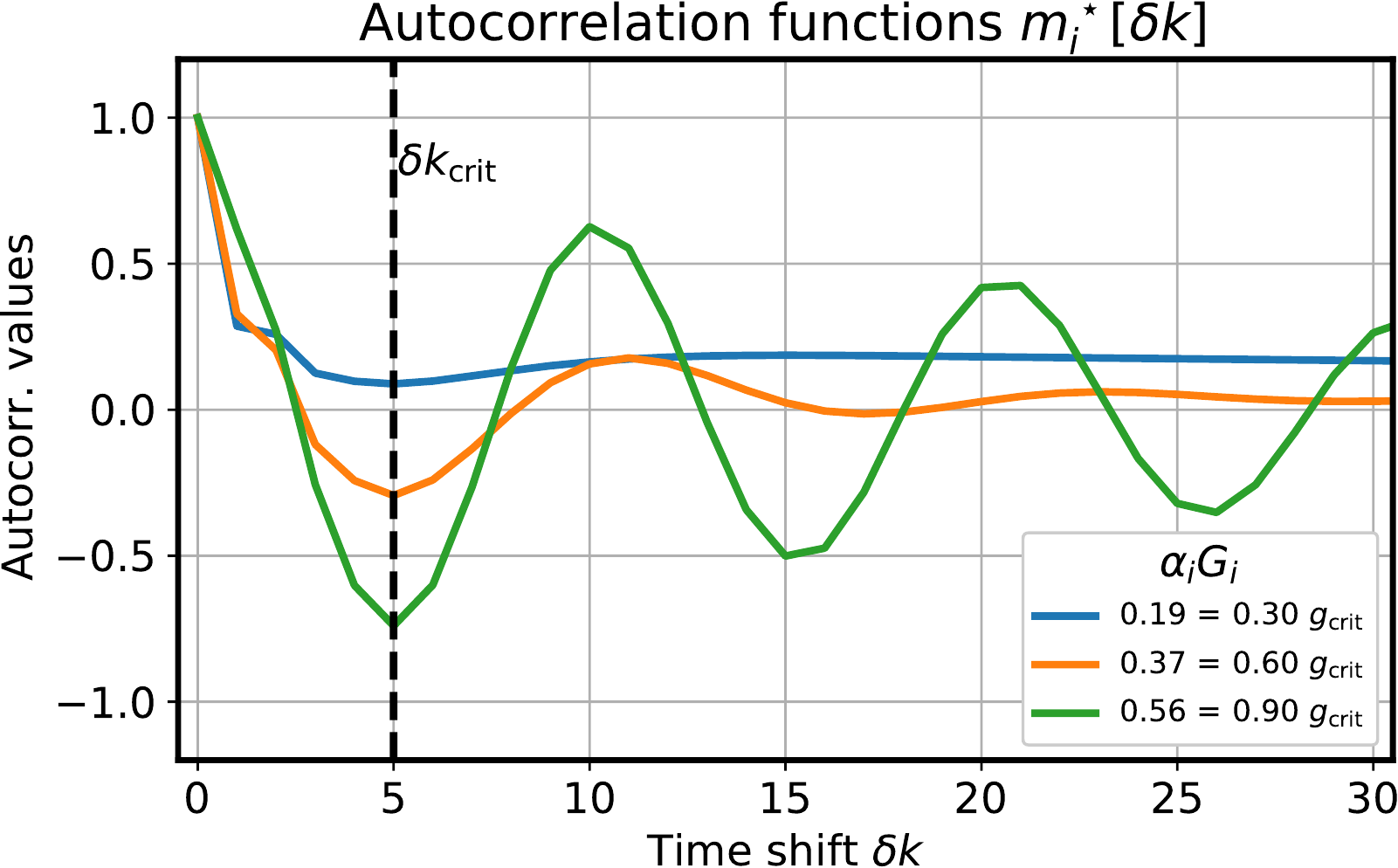}
	\else
		\includegraphics[width=.98\columnwidth]{3_ACplots.pdf}
	\fi
	\caption{
		Autocorrelation functions $m_i^\ast[\dk]$ corresponding to the measurement PSDs $|\hat{m}_i|^2$ shown on Fig.~\ref{fig:3_spectralOutput}.
		$\dkc = 5$~frames for $\dfrac{\tau}{T} = 2$.
	}
	\label{fig:3_ACplots}
\end{figure}

Instead, we propose to use the autocorrelation (AC) of the modal measurements, noted $m_i^\ast[\dk]$,
as it provides another indirect measure of $\alpha_i G_i$.
For reference, the AC curves $m_i^\ast[\dk]$ at small time shifts $\dk$ corresponding to the measurement spectra shown on the bottom panel of Fig.~\ref{fig:3_spectralOutput} are shown on Fig.~\ref{fig:3_ACplots}.
As $\alpha_iG_i$ increases and as the loop response approaches divergence, an oscillation of half-period nearing
\begin{equation}
	\dkc = \dfrac{f_S}{2\fc} = 2\dfrac{\tau}{T}+1\text{ frames}
\end{equation}
is superimposed over the typically wider bell-shaped AC curve; this oscillation is the correspondence in the AC domain of the resonance peak, growing in amplitude and converging to $\fc$.

In particular, the first minimum of the AC curve, for a time shift of $\dkc$, is a strongly marked signal that allows for a measurement of the amount of resonance in the loop, with a minimal latency of $\dkc$ frames.
This value is the one metric that is monitored by CLOSE to register implicitly the value of the loop gain $\alpha_i G_i$:
\begin{equation}
m_i^\ast[\dkc] = \dfrac{\displaystyle\sum_k m_i[k]m_i[k-\dkc]}{\displaystyle\sum_k m_i[k]^2},
\end{equation}
which is also, as we will see in Sect.~\ref{sec:4_impl}, convenient to track with simple real-time estimators.
Decreasing $\alpha_i G_i$ reduces the amplitude of the oscillation over the AC function, which tends to increase the value of $m_i^\ast[\dkc]$; and the opposite occurs when increasing $\alpha_iG_i$ (within the stability limits).

The relation between $\alpha_iG_i$ and $m_i^\ast[\dkc]$ being monotonic, one can act on $G_i$ in order to \emph{lock} the $\dkc$-correlation value onto a steady-state solution
\begin{equation}
\label{eq:3_steadyState}
m_i^\ast[\dkc] = r \in [-1, 1],
\end{equation}
where we call $r$ the \emph{setpoint} value.
Without additional knowledge, the value of $r$ ought to be adjusted (or defined per-mode) as to fit a performance-maximizing criterion in all useful situations the AO would face, and across the complete range of the effective modal S/N.
The automatic driving of $G_i$ to satisfy Eq.~\ref{eq:3_steadyState} is of course particularly convenient if and only if a unique, or a small number of setpoint values may be found.
%, which is the only free remaining parameter of our AO loop model.

The $\dkc$-correlation value, and the value $r$ we target to lock it onto, are empirically representative of the spectral energy ratio between the low-frequency atmospheric rejection residual, and the overshoot peak near $\fc$, as can be seen on Figs.~\ref{fig:3_spectralOutput} and~\ref{fig:3_ACplots}.

Generally speaking, using higher values for $r$ will impose a more cautious and robust steady-state control solution with a lower loop gain - with a strong, near unit correlation at $\dkc$ separation.
And smaller values will lead to more aggressive loop behaviors, possibly reaching nearly-divergent transfer functions, but with a maximized rejection of the low frequency components.
As $\alpha_i G_i$ approaches $g_\mathrm{crit}$, $m_i^\ast[\dkc]$ goes towards -1, and the output of the system is a slowly dampened sinusoid of period $2\dkc$ frames.

Furthermore, with the condition of Eq.~\ref{eq:3_steadyState} satisfied, CLOSE enforces a transfer function constraint that is independent of the sensitivity reduction $\alpha_i$ of the WFS.
In other words, it automatically compensates for the optical gain effect using modal compensation coefficients.
%Therefore, it provides a go-around strategy regarding the impossibility to apply to the PWFS (or other systems compatible with the description discussed in Sect.~\ref{sec:2_model}) some methods \citep{Gendron1994Astronomical, Dessenne1998Optimization} based on numerical estimations of transfer functions that are restricted to linear systems without unknowns.

\subsection{Correlation-locking and minimum variance: semi-analytical solutions}
\label{sec:3_3_CLOSE_MV_comparisons}

We have introduced the concept of a correlation-locking condition to adjust the controller gains and achieve a given control transfer function independently of the sensitivity reduction.
We propose to compare the solutions obeying Eq.~\ref{eq:3_steadyState} to a minimum variance (MV) modal integrator, verifying Eq.~\ref{eq:2_3_PSDVariance}.

\begin{figure}
	\centering
	\ifreferee
		\includegraphics[width=0.7\columnwidth]{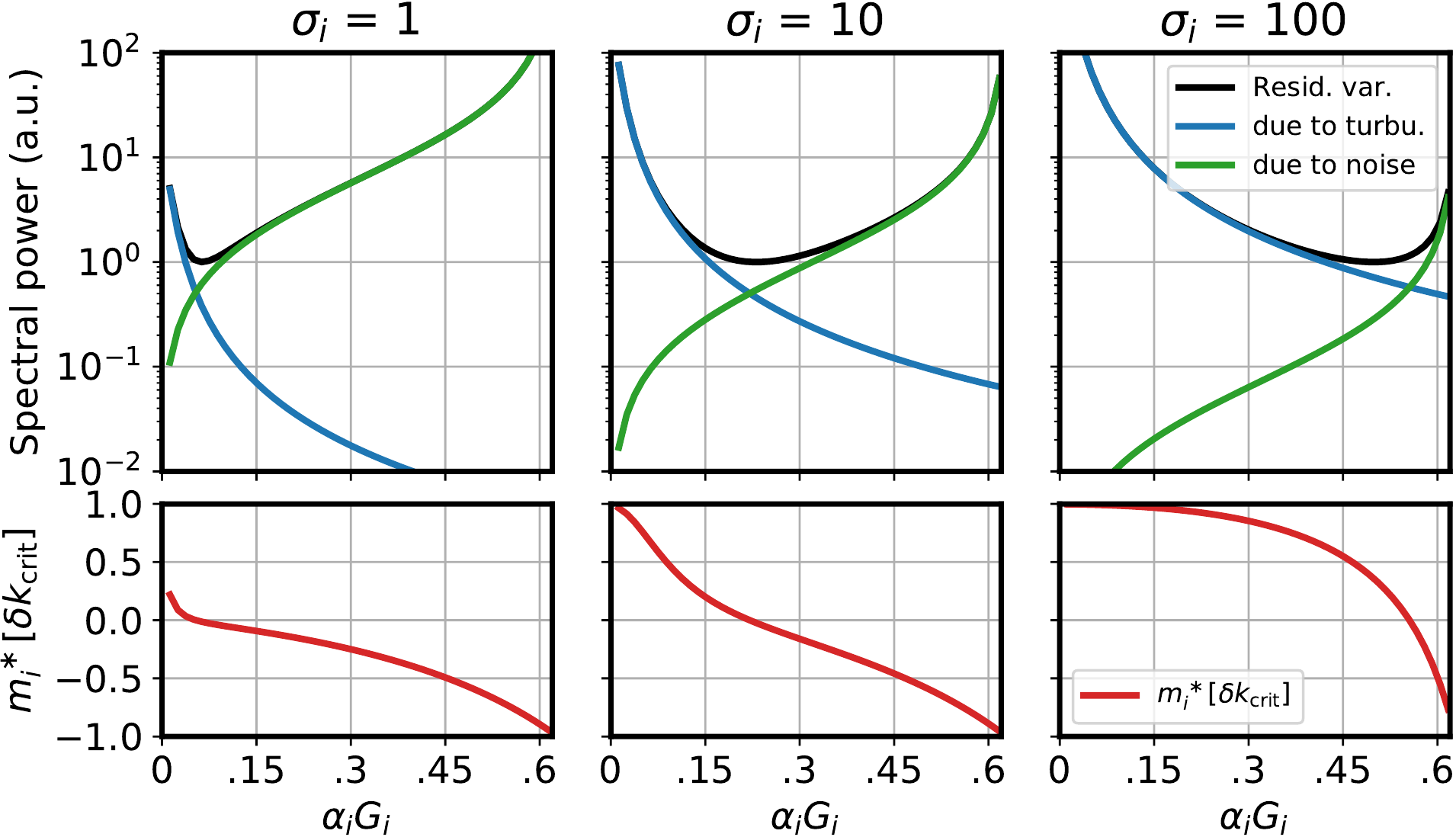}
	\else
		\includegraphics[width=\columnwidth]{3_moreAC.pdf}
	\fi
	\caption{%
		For effective S/N $\sigma_i$ = 1, 10 and 100, closed-loop residual variances (top panels) and resulting $\dkc$-autocorrelation of the measurements (bottom panels), as a function of the loop gain $\alpha_i G_i$ from 0 up to $g_\mathrm{crit}$.
		The latency is taken as $\tau = 2.T$.
		Black lines - total residual variance $\int_f \left|\hat{\phi}_\mathrm{Res}^i(f)\right|^2\dd f$ - normalized to 1 at its minimum; blue: variance induced by turbulence $\int_f \left|h^i_{\mathrm{Atm}\longrightarrow\mathrm{Res}}\cdot\hat{\phi}_\mathrm{Atm}^i(f)\right|^2\dd f$;
		and green: variance induced by noise $\int_f \left|h^i_{\mathrm{n}\longrightarrow\mathrm{Res}}\cdot\frac{1}{\sigma_i}\right|^2\dd f$.
	}
	\label{fig:3_var_vs_ac}
\end{figure}

For the example developed in Figs.~\ref{fig:3_spectralOutput} and~\ref{fig:3_ACplots}, the gain $\alpha_i G_i$ minimizing the variance of the wavefront residual is $g_\mathrm{MV}\approx~0.231$.
At this gain value, we obtain $m^\ast_i[\dkc] = -0.025$.
This numerical observation hints towards pursuing $r=0$ as a candidate setpoint.

The computations are easily generalized for values other than the sensitivity-corrected S/N $\sigma_i = 10$ used for the examples.
We show on Fig.~\ref{fig:3_var_vs_ac} the variations of $m_i^\ast[\dkc]$ and of the residual variance (Eq.~\ref{eq:2_3_PSDVariance}) with $\alpha_i G_i$, for S/N values $\sigma_i =$ 1, 10, and 100.
We also plot the two components of the variance due to the turbulence residual and the noise propagation, i.e., the two components that the CLOSE servo-loop seeks to balance optimally.
The observed variations of $m_i^\ast[\dkc]$ confirm its monotonicity with the gain, as well as the relationship inferred in Sect.~\ref{sec:3_2_CLOSE}: a negative observable $m_i^\ast[\dkc]$ (or setpoint $r$) relates to a high gain loop, and a positive $m_i^\ast[\dkc]$ to a low gain loop.
For all three S/N, we observe on Fig.~\ref{fig:3_var_vs_ac} the approximate match between the gain yielding minimum variance and the intercept $m_i^\ast[\dkc] = 0$.
This further establishes pursuing $r=0$ as a special value usable in a variety of situations.

\begin{figure}[t]
	\centering
	\ifreferee
	\includegraphics[width=.65\columnwidth]{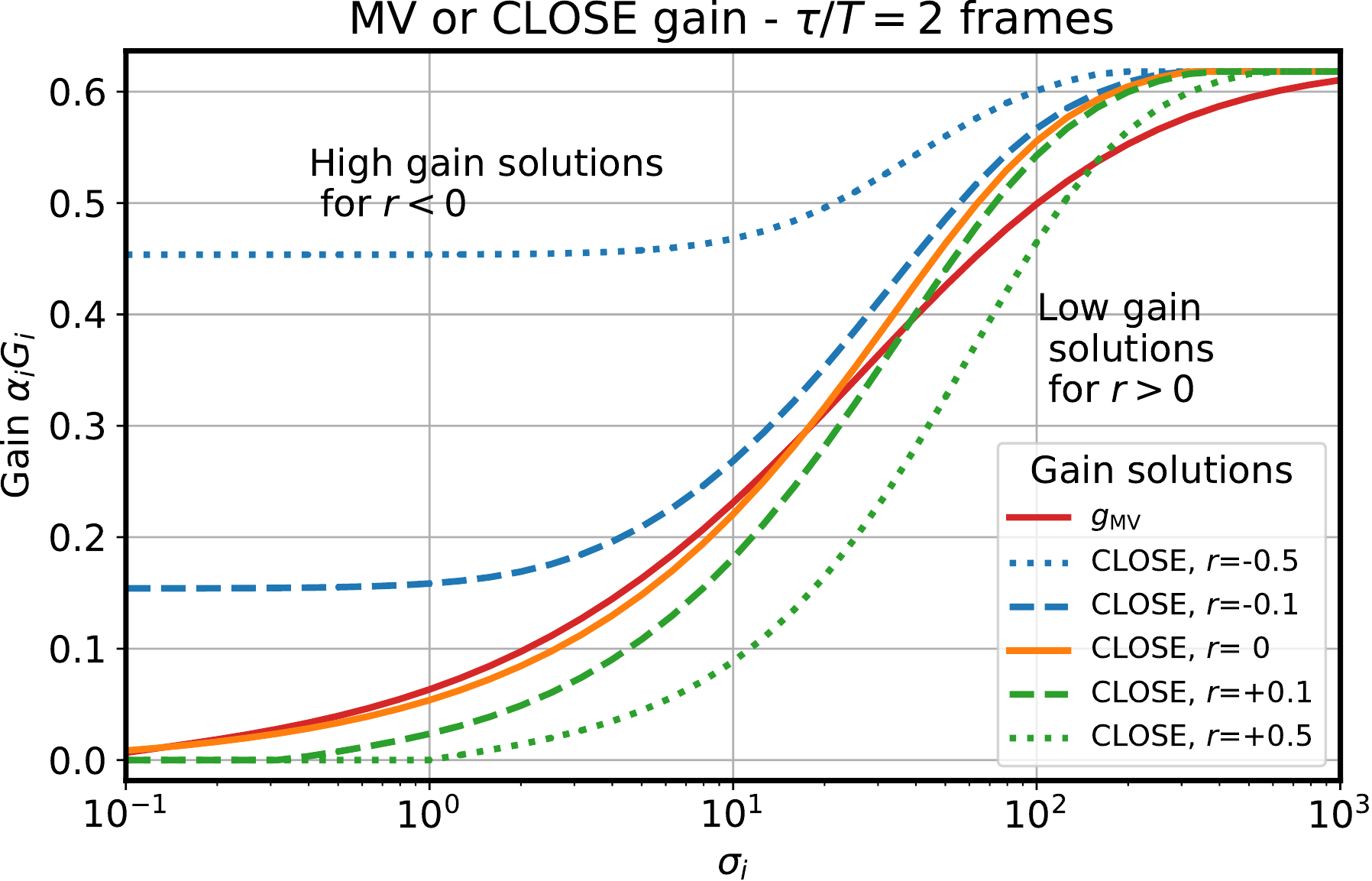}
	\else
	\includegraphics[width=.98\columnwidth]{3_CLOSE_MV_lat2.pdf}
	\fi
	\caption{%
		Minimum variance gain $g_\mathrm{MV}$ and solutions found using CLOSE with five different setpoint values, depending on the sensitivity corrected S/N $\sigma_i$.
		These computations are performed using the modal input spectrum of Eq.~\ref{eq:2_3_turbSpectrum}, with $f_i=1$~Hz.
	}
	\label{fig:3_CLOSE_MV_comp}
\end{figure}

Generalizing further these analyses, we show on Fig.~\ref{fig:3_CLOSE_MV_comp} the values of $\alpha_i G_i$ resulting in setpoints $r$ of -0.5, -0.1, 0, 0.1 and 0.5, as well as the MV solution, for effective S/N values $10^{-1}$ to $10^{4}$.
The gain $g_\mathrm{MV}$ minimizing the variance for a given S/N is found --following \citet{Gendron1994Astronomical}-- by numerically minimizing the joint Eqs.~\ref{eq:2_HatmToRes} and~\ref{eq:2_HnToRes}.
CLOSE solutions are found by numerically solving $m_i^\ast[\dkc] = r$ for $\alpha_i G_i$, through Eqs.~\ref{eq:2_HatmToM} and~\ref{eq:2_HnToM}.
Additional data for $\tau = 0$ and $\tau = T$ is presented in Appendix~\ref{appdx:B_CLOSE_MV}.
%\red{More extensive sims may be shown in an appendix if that's of any relevance.}

First, is should be noted how the correlation-locked solution for an empirical $r=0$ is such a remarkable close match to the MV solution $g_\mathrm{MV}$.
The discrepancy reaches --at worst-- respectively 20\% and 15\% of $g_\mathrm{MV}$ in the two knees of the curve, near $\sigma_i=3$ and $\sigma_i=100$.
We also confirm the statement made earlier: positive setpoints lead to lower values of $\alpha_i G_i$ for a given $\sigma_i$, while negative ones lead to a higher gain value.
Interestingly, negative setpoints impose a ``gain floor'', even when the S/N is extremely poor; indeed, to achieve $r<0$ over essentially white noise, the controller forcefully increases the gain as to introduce a detectable oscillation of the amplified noise.
Overall, $r=0$ seems to provide an approximate near-MV solution over most of the $\sigma_i$ range, with some room for improvement using a combination of $r>0$ values for the higher end of the S/N range.
The exact discrepancy between the $g_\mathrm{MV}$ and the CLOSE solution at null setpoint of course depends on the actual spectrum of turbulence, in particular the cutoff frequency $f_i$.

Altogether, these semi-analytical simulations show that reaching the CLOSE steady state solution using $r=0$ can provide a near minimum variance solution throughout the entire range of $\sigma_i$ - thus making it a relevant control technique for all the controlled DM modes, for any choice of guide star magnitude, for any amount of input turbulence, as well as for any sensitivity reduction $\alpha_i$, given that all these parameters merely factor in the computation of $\sigma_i$.
With a unique setpoint value, semi-analytical computations vouch for a simple criterion that would enable fully automatic (almost) minimum variance integral control, even for sensors with poorly modeled sensitivity variations.

We will come to validating these claims using end-to-end numerical simulations in Sect.~\ref{sec:5_results}; we present in the next section how we propose to practically implement the steady state equations presented so far.
%\clearpage
% We might have to split this section, it's gonna be FAT

%  ___                 _                           _        _   _             
% |_ _|_ __ ___  _ __ | | ___ _ __ ___   ___ _ __ | |_ __ _| |_(_) ___  _ __  
%  | || '_ ` _ \| '_ \| |/ _ \ '_ ` _ \ / _ \ '_ \| __/ _` | __| |/ _ \| '_ \ 
%  | || | | | | | |_) | |  __/ | | | | |  __/ | | | || (_| | |_| | (_) | | | |
% |___|_| |_| |_| .__/|_|\___|_| |_| |_|\___|_| |_|\__\__,_|\__|_|\___/|_| |_|
%               |_|                                                           
\section{Practical implementation}
\label{sec:4_impl}

\subsection{Real-time}
\label{sec:4_1_realtime}

Modal WFS telemetry $m_i[k]$ is obtained at each time step $k$, as the product of the WFS output by the modal reconstructor $\mathbf{Rec}$.
Two online autocovariance estimators are built from the $m_i[k]$ using discrete integrators:
\begin{align}
\label{eq:4_ACEstim}
N_i^0[k]\ \, &= p\ m_i[k]^2 + (1-p)N_i^0[k-1]
\nonumber \\
N_i^{\mathrm{crit}}[k] &= p\ m_i[k]m_i[k-\dkc] + (1-p)N_i^{\mathrm{crit}}[k-1],
\end{align}
where $k$ is the time step index, and $p \in [0, 1]$ a smoothing parameter.
$N_i^0$ and $N_i^{\mathrm{crit}}$ are thus lowpass time-filtered estimates such that $N_i^0$ tracks the series variance $\mathrm{Var}_k(m_i[k])$, and their ratio $N_i^{\mathrm{crit}} / N_i^0$ tracks the critical autocorrelation $m_i^\ast[\dkc]$.
After an empirical optimization of the parameter $p$, we opted for fast integrators with $p = 0.3$ for all simulations presented in Sect.~\ref{sec:5_results}.
One objective achieved through this smoothing parameter is simply bridging the gap of $\dkc$ frames between the joint estimations of the variance and of $m_i^\ast[\dkc]$.

After the AC estimation, the modal gains $G_i$ are updated using multiplicative increments as follows:
\begin{equation}
\label{eq:4_GUpdate}
G_i[k] = G_i[k-1] \times \left[ 1 + q^\pm \left( \dfrac{N_i^{\mathrm{crit}}[k]}{N_i^0[k]} - r \right) \right].
\end{equation}
The $r$ parameter being the loop setpoint as defined in section~\ref{sec:3_closeTheory}.

The $q^\pm$ learning factor may encompass two different values, with either $q^+$ and $q^-$ used depending on the sign of
$$\dfrac{N_i^{\mathrm{crit}}[k]}{N_i^0[k]} - r,$$
$q^+$ being used for $G_i$ increases in Eq.~\ref{eq:4_GUpdate}, and $q^-$ for decreases.
This asymmetry is kept as an option as to make the algorithm more reactive to overshooting transients (using $q^- > q^+$), as compared to tracking gain increases due to a transfer function deemed too slow (using $q^+$).

This asymmetric tracking is not used for results shown in this paper but was used in preliminary numerical simulations (and throughout \citealp{Deo2019CLOSE}); it allows to maintain stability at higher $q^{\pm}$ values.
It is to be noted however that using $q^+ \neq q^-$ induces a bias in the mean value achieved for $m_i^\ast[\dkc]$, then differing from $r$.
We empirically observed that e.g. for $q^- = 5q^+$, $r \in [- 0.2, - 0.1]$ should be used to retrieve the performance seen with $r = 0$ in the $q^+ = q^-$ case.

The $q^\pm$ learning factors determine the time constants associated with the convergence and tracking ability of the CLOSE servo-loop.
We infer that for a real AO system, $q^\pm$ values in the range of $10^{-3}-10^{-4}$ should be used (assuming 500~Hz frequency), hence providing typical time constants in the $2-20~$seconds range.
The ideal choice of $q^\pm$ will probably remain dependent on the system, and will certainly require some adjustments accounting for robustness and responsiveness to variations of turbulence conditions, vibrations, or other transient events.

While the theoretical derivations were most accurate using $r=0$, we do not exclude that for each system, some sort of empirical tweaking of $r$ may be necessary to either privilege consistent stability or aggressive rejection.

\subsection{Computational strain}
\label{sec:4_2_computeStrain}

Implementing CLOSE in a real-time fashion is of course expected to increase the AO loop computational requirements.
While the AC estimations and gain updates themselves (Eqs.~\ref{eq:4_ACEstim} and \ref{eq:4_GUpdate}) are negligible compared to the matrix-vector multiplication (MVM), having the $m_i[k]$ available in real-time requires to do the reconstruction in two successive MVM steps.
The first MVM converts WFS measurements to modal values, with a computational burden nearly identical to the usual overall MVM from measurements to DM commands.
The second step computes DM increments from modal values, with a nearly square matrix of size the number of actuators.
While this two-step operation is not universal, it is worth noting it is used routinely on some instruments~(e.g., \citealp{Guyon2018CACAO}).
While this two-step technique may become a computational bottleneck, in particular for high contrast systems on ELTs, it is currently baselined for the control of the deformable M4 on ESO's ELT~\citep{Bonnet2019Wavefront}.

For a typical PWFS AO system with some spatial oversampling (the pixel projected size is smaller than the DM pitch), the number of pixels read out is typically 5 to 6 times the number of actuators --there being 4 pupil-like images, times the square of the oversampling factor.
The number of outputs of the PWFS is therefore 2.5 to 6 times the number of modes, depending whether slopes-map preprocessing is used or not. The algorithmic interest and computing cost of skipping the preprocessing is discussed in, e.g.,~\citet{Deo2018Assessing}.
While the first MVM execution time depends on the WFS output dimensionality, the second MVM depends only on the number of controlled modes, with a smaller, albeit not negligible, computational footprint.
As an example, we measured the RTC computation time for the AO simulation setup used across Sect.~\ref{sec:5_results} (see Table~\ref{tab:5_paramTable}): 881.6 $\pm$ 4.2 \textmu s using a single MVM, against 1035.6 $\pm$ 3.6 \textmu s using two cascaded MVMs. These timings where achieved using a single Nvidia Tesla P100 graphical processor.

\subsection{Offline implementation}
\label{sec:4_3_offline}

If the real time computer (RTC) software cannot be altered on an existing system, or if the additional strain is not acceptable within the RTC specifications, CLOSE can be implemented in a block-wise manner.
All estimators, gain updates, and command matrix updates are performed in offline time, certainly in another process and preferably on another machine, over batches of contiguously recorded measurements.
This buffered strategy enables to deploy CLOSE on nearly any existing AO system that reliably provides its WFS telemetry without excessive delays.

A time-continuous buffer of $K$ WFS measurements is forwarded to the CLOSE process, which turns them into modal measurements $m_i[0],\ ...,\ m_i[K-1]$ using the modal reconstructor $\mathbf{Rec}$.
For each mode, the AC estimators of Eq.~\ref{eq:4_ACEstim} are replaced by the direct computation of the normalized $\dkc$-shifted AC term over the telemetry buffer:
\begin{align}
\label{eq:4_blockEstim}
N_i^\mathrm{block} = \dfrac{\dfrac{1}{K-\dkc}\displaystyle \sum_{k=\dkc}^{K-1} m_i[k]m_i[k-\dkc]
}{
	\dfrac{1}{K}\displaystyle \sum_{k=0}^{K-1} m_i[k]^2}.
\end{align}

The gain update equation can then be performed:
\begin{equation}
G_i[\mathrm{new}] = G_i[\mathrm{previous}] \times \left[1 + q^\pm \left( N_i^\mathrm{block} - r \right) \right],
\end{equation}
using $q^\pm$ factors adjusted for the longer integration time and the increased SNR on AC estimation; typically $q^\pm$ ought to be larger by a factor $\sqrt{K}$ for a dynamical effect comparable to the real-time implementation.
The new command matrix can then be computed accounting for the new $G_i$ values, and when all side-tracked computations are finished, can be set into the RTC.

%  ____                 _ _       
% |  _ \ ___  ___ _   _| | |_ ___ 
% | |_) / _ \/ __| | | | | __/ __|
% |  _ <  __/\__ \ |_| | | |_\__ \
% |_| \_\___||___/\__,_|_|\__|___/
\section{Numerical simulations results}
\label{sec:5_results}

This section covers some end-to-end numerical simulations demonstrating the performance achieved with CLOSE when applied to the MICADO SCAO design \citep{Clenet2019MICADO, Vidal2019Analysis}.
The main parameters of the system and the simulations are summarized in Table~\ref{tab:5_paramTable}.
All simulations were performed using the COMPASS platform \citep{Ferreira2018COMPASS}.

In Sect.~\ref{sec:5_1_gainconvergence}, we analyze the convergence of modal gains when bootstrapping the AO loop.
In Sect.~\ref{sec:5_2_stationary}, we verify the steady-state performance achieved using $r=0$, for various S/N levels.
This is expanded in Sect.~\ref{sec:5_3_r_validation}, exploring the impact of different setpoint values.
Finally, we show in Sect.~\ref{sec:5_4_transients} the dynamic behavior of CLOSE when exposed to sudden changes in conditions.
Throughout the following sections, seeing conditions are referred to using the Fried parameter $r_0$, always given at 500~nm; and guide star brightnesses are identified by the R-band magnitude ($M_R$), related to the photon flux per the zero-point and system throughput given in Table~\ref{tab:5_paramTable}.
AO performance is most often measured in terms of H-band long exposure (LE) Strehl ratios (SR), computed from simulated monochromatic point spread functions at 1\,650~nm.
SR comparisons given in \% are always in percentage point units, never relative variations.

%Table moved to force columns to behave.
\begin{table}[t]
	\centering
	\caption{%
		AO numerical simulation parameters.
		Values within brackets indicate probing ranges for various simulations reported across Sect.~\ref{sec:5_results}.
		Values preceding brackets indicate the default value for the parameter.
	}
	\label{tab:5_paramTable}
	\renewcommand{\arraystretch}{1.1}
	\begin{tabular}{ll}
		\hline\hline
		\multicolumn{2}{c}{Numerical simulation configuration}\\
		\hline
		\multirow{2}{*}{Telescope} %
		& $D$ = 39~m diameter\\ 
		& ELT pupil model (no spider arms)\\
		\hline
		\multirow{3}{*}{Turbulence} %
		& von Kármán, ground layer only\\
		& $r_0$ at 500~nm: $[8.9-21.5]$~cm\\
		& L$_0$ = 25~m\\
		& speed: 10~m.s$^{-1}$ $[10-40]$\\
		\hline
		\multirow{2}{*}{Guiding} %
		& On-axis natural guide star\\
		& Zero point: $2.6 \times 10^{10}$~ph.s$^{-1}$.m$^{-2}$\\
		& Magnitude $M_R$: [0-18]\\
		\hline
		\multirow{4}{*}{DMs} %
		& Tip-tilt mirror\\
		& Hexagonal M4 model pattern\\
		& $\quad$Pitch of 54~cm, coupling of 0.24\\
		& $\quad$4\,302 controlled actuators\\
		\hline
		PWFS & \\
		\multirow{2}{*}{$\quad$Subapertures} %
		& 92$\times$92 -- pixel size 42~cm.\\
		& 24\,080 pixels used for control\tablefootmark{(a)}\\
		$\quad$Wavelength & Monochromatic, 658~nm\\
		$\quad$Throughput & 0.28 (including quantum efficiency)\\
		$\quad$Modulation & Circular, 4~$\frac{\lambda}{D}$ radius\\
		$\quad$Readout noise & 0.3~$e^{-}$\\
		\hline
		RTC controller & \\
		$\quad$Loop rate & $f_S=500$~Hz\\
		$\quad$Latency & $\tau = 2T = 4$~ms \tablefootmark{(b)}\\
		\multirow{2}{*}{$\quad$Method} %
		& Two step modal linear integrator\\
		& Pixels~$\longrightarrow [m_i] \times [G_i] \longrightarrow$~actuators\\
		$\quad$Basis & DM Karhunen-Loève basis\tablefootmark{(c)}\\
		\hline
		\multirow{2}{*}{CLOSE} %
		& Real-time implementation\\
		& $p = 0.3$; $q^\pm = 10^{-3}$; $\dkc=5$\\
		& $r=0.$ [$-0.5, 0.5$].
	\end{tabular}
	\tablefoot{
		\tablefoottext{a}{Bypassing the slope maps computation, as per \citet{Guyon2005Limits, Clergeon2014Etude, Deo2018Assessing}.}
		\tablefoottext{b}{Extended to $\tau$ in [0, $T$, $2T$] in Appendix~\ref{appdx:A_r_other}.}
		\tablefoottext{c}{\citet{Ferreira2018Adaptive}.}
	}
\end{table}

\subsection{Gain convergence upon closing the loop}
\label{sec:5_1_gainconvergence}

\begin{figure}[t]
	\centering
	\ifreferee
	\includegraphics[width=0.65\columnwidth]{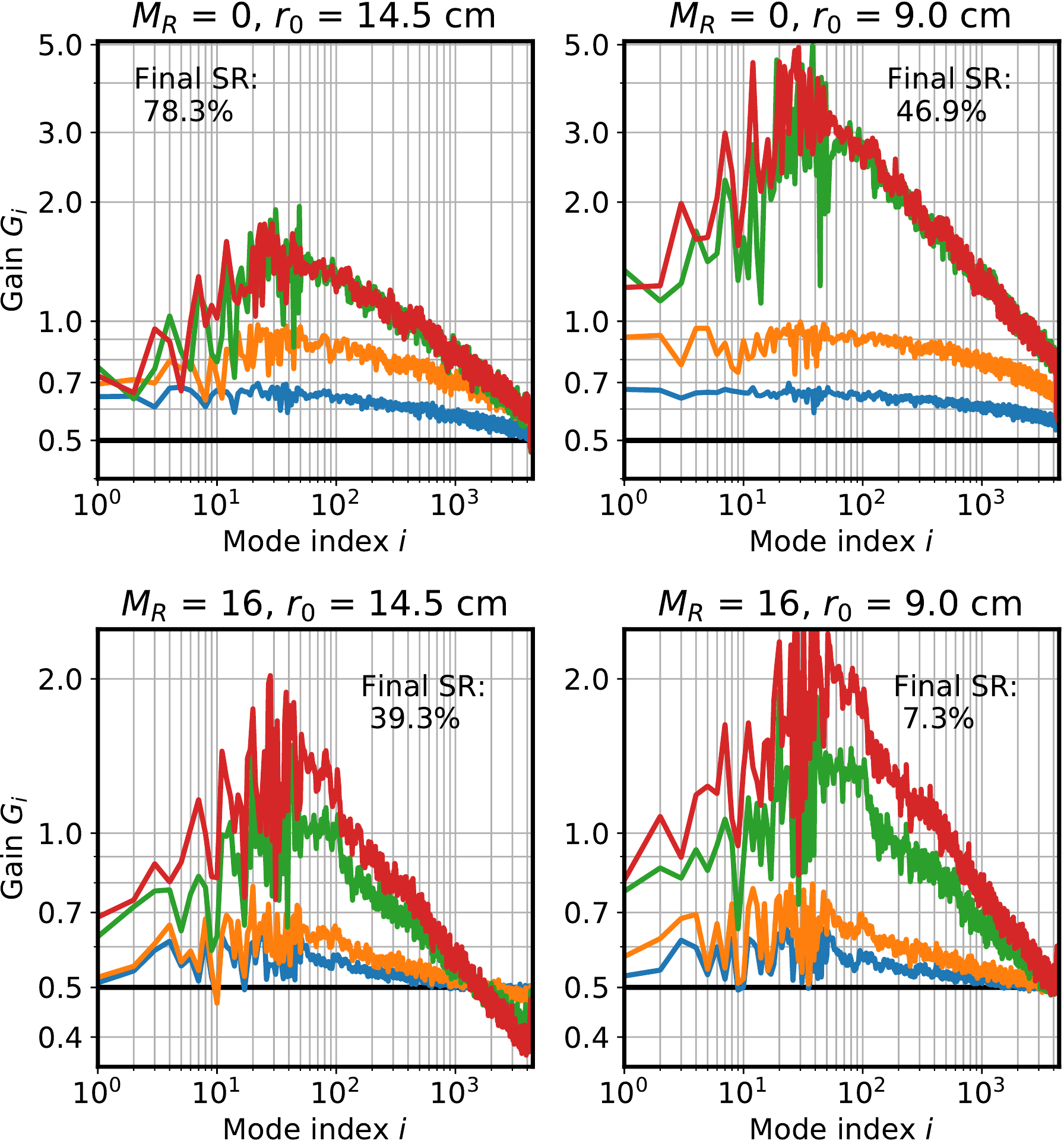}\\[.4em]
	\includegraphics[width=0.5\columnwidth]{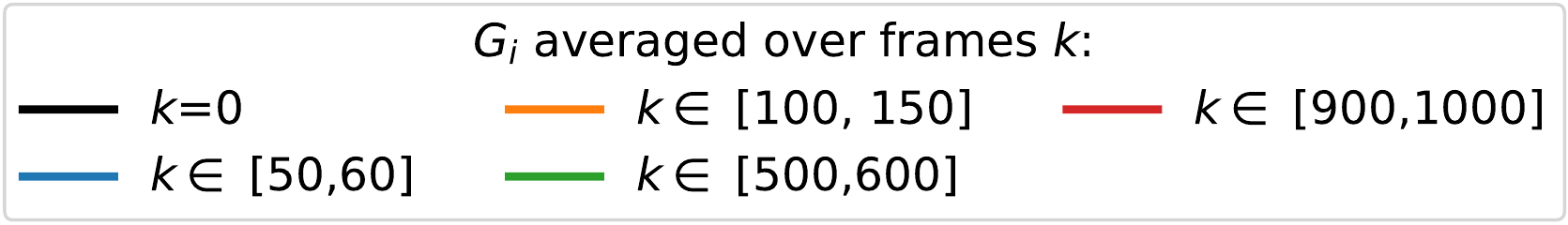}
	\else
	\makebox[\columnwidth][c]{% % If this figure ends up if the left column
		\includegraphics[width=1.05\columnwidth]{5_bootstrap.pdf}%
	}
	\\[.4em]
	\includegraphics[width=0.75\columnwidth]{5_bootstrapLegend.pdf}
	\fi
	\caption{
		Convergence of CLOSE gains on the 2.0~sec following closing of the AO loop, for guide stars of $M_R=0$ and 16 and atmospheric $r_0$ of 14.5 and 9.0~cm.
		All $G_i$ are initialized to 0.5 at start-up (blue line).
		$G_i$ values are shown as averaged over the time windows given in the legend. Curves are smoothed along the $i$ index for clarity.
		Final SRs are given in H-band and computed from the cumulative exposure over the last 200~msec ($k \in [900, 1000]$).
	}
	\label{fig:5_bootstrap}
\end{figure}

We first investigate the dynamics of the modal gains upon closing the AO loop with CLOSE enabled.
These simulations are all initialized identically, regardless of $r_0$ or $M_R$; we opt for the starting value $G_i[k\!=\!0] = 0.5$ for all modes.
With the system latency $\tau = 2T$ simulated, the critical gain value is $g_\mathrm{crit} \approx 0.61$; given that the sensitivity reduction $\alpha_i$ is always smaller than 1, this ensures that the loop is closed with a comfortable stability margin.
From these initial 0.5 values, the $G_i$ are driven by CLOSE to their steady-state values, accounting both for non-linearity compensation and temporal variance minimization.

We show on Fig.~\ref{fig:5_bootstrap} the temporal averages of the 4\,301 modal gains $G_i$, for several time windows within the first two seconds after the AO loop is closed.
These simulations are performed for four different cases, with $r_0$ of 14.5 and 9.0~cm, and guide stars of brightnesses $M_R = 0$ and $M_R = 16$.

For the bright cases, a steady state is reached by frame $k\!\approx\!500$, i.e., within one second. The process is slightly slower for the $M_R=16$ cases, with a continued convergence of the $G_i$ between frames $k=500$ and $k=1000$.
Simulations at $M_R = 0$ are essentially performed with an infinite S/N. As such, the dynamical gains $\alpha_i G_i$ evolve nearby the maximum stability value $g_\mathrm{crit}$, and the $G_i$ coefficients reached in steady-state are essentially reflecting the inverse $\alpha_i^{-1}$ of the PWFS sensitivity reduction.
The $G_i$ curves reached at the end of convergence in $M_R=0$ cases are in good accordance with the abacuses presented in previous work \citep{Deo2018Modal, Deo2019Telescope} --obtained by directly measuring $\alpha_i$ sensitivity reductions on static turbulence screens--, with $\alpha_i$ typically depending on mode index $i$ as: decreasing up to mode 30, which contains spatial frequencies corresponding to the modulation radius, then increasing again roughly as a power law up to the highest order modes.

\begin{figure}[t]
	\centering
	\ifreferee
	\includegraphics[width=0.7\columnwidth]{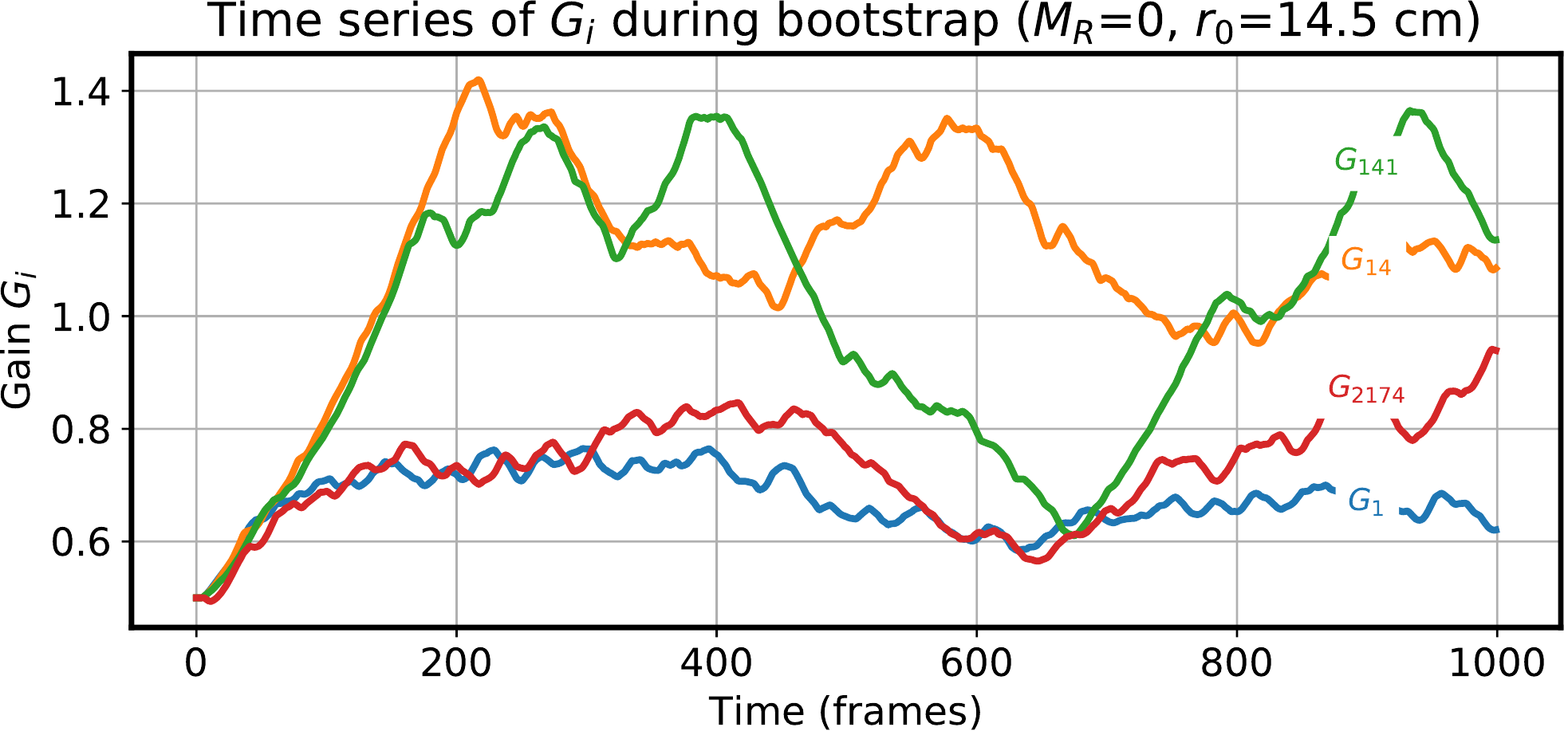}
	\else
	\includegraphics[width=0.99\columnwidth]{5_bootstrapTimeSeries.pdf}
	\fi
	\caption{
		Time series of CLOSE gains for four select modes on the 2.0~sec following closing of the AO loop, for the simulation $r_0 = 14.5$~cm, $M_R=0$ shown on Fig.~\ref{fig:5_bootstrap}.
	}
	\label{fig:5_bootstrapTimeSeries}
\end{figure}

Some more insight on the bootstrapping of CLOSE can be gained by inspecting temporal series of modal gains, as shown on Fig.~\ref{fig:5_bootstrapTimeSeries}; these series correspond to four modes, and are the same data set that on Fig.~\ref{fig:5_bootstrap} for $r_0=14.5$~cm, $M_R=0$.
While the AO loop bootstraps in only a few frames, the convergence of CLOSE takes a longer time, and induces some modal gain oscillations after the initial ramp-up.
The amplitude and time constants of these oscillations certainly depends on a number of parameters, and importantly on the learning factor $q^\pm$.
On Fig.~\ref{fig:5_bootstrapTimeSeries}, various periods are observed --from $\approx$ 1~sec down to smaller oscillations typically every ~30~frames (60~msec).

The randomness of the turbulence screens is certainly a factor in continuous variations of the $G_i$, and the oscillations continue even as the AO integrator and CLOSE reach steady states, beyond 1-2~secs. of runtime, albeit with little impact on the correction quality.
Figs.~\ref{fig:5_seeingBurst} and~\ref{fig:5_cloudBurst} will show complementary data to Figs.~\ref{fig:5_bootstrap} and~\ref{fig:5_bootstrapTimeSeries}, including longer time series and Strehl ratios.

Furthermore, the physical nature of the PWFS is certainly a contributor to the $G_i$ fluctuations, as compared to a purely linear sensor.
In case that the $G_i$ reach values that are too high, or if we had started from $G_i$ values higher than the steady-state ultimately reached,
the optical gain phenomenon itself helps in maintaining stable, sub-optimal control states while the controller performs the slow gain decrease.
To summarize this effect, reduced atmospheric residuals (e.g., bootstrapping, or improving conditions) induce an increase in WFS sensitivity; with the gains $G_i$ being overset, transfer functions become highly resonant or temporarily unstable; but in turn, the added wavefront residual --from transitory divergence of the loop or noise oscillations-- induces a reduction in the PWFS sensitivity.
This regime is progressively stabilized as the adaptive filtering eventually adapts the command law to the ongoing conditions and reaches a steady state regime.
These transitory effects will be further discussed in Sect.~\ref{sec:5_4_transients}.

When comparing from Fig.~\ref{fig:5_bootstrap} the behavior between $M_R=0$ and $M_R = 16$ cases, one can observe the effect of the implicit optimization of the transfer function, with steady-state gain values dampened by typically 20-50\% in the dim case relative to the bright one, depending on the mode number and the $r_0$.
Altogether, results presented on Fig.~\ref{fig:5_bootstrap} tend to validate that-- without any priors and regardless of the PWFS sensitivity reduction-- CLOSE successfully drives the modal integrator gains to convergence, across a 1 to 2~second period.

\subsection{Performance in stationary conditions for $r=0$}
\label{sec:5_2_stationary}

\begin{figure}[t!]
	\centering
	\ifreferee
		\includegraphics[width=.7\columnwidth]{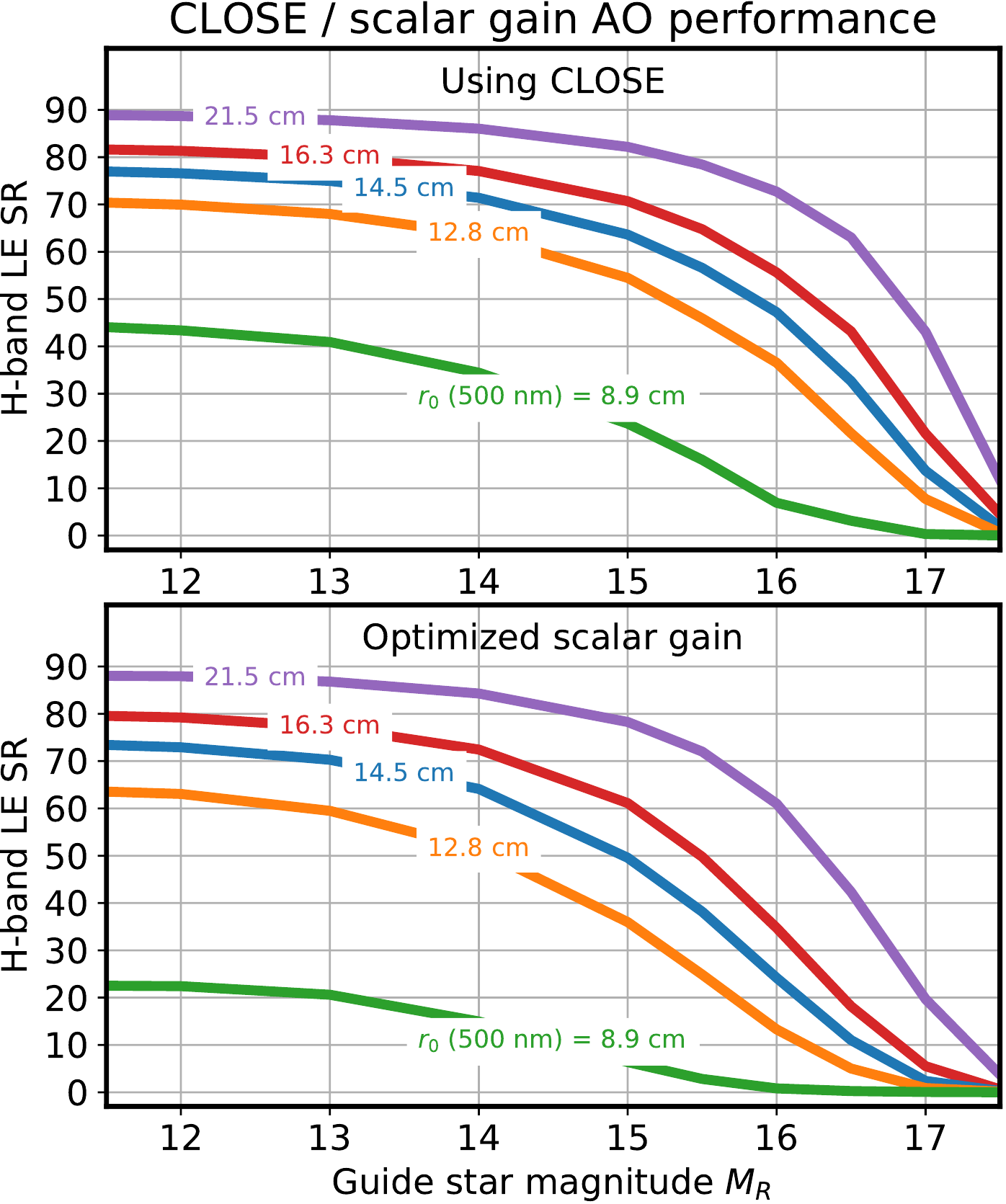}
	\else
		\includegraphics[width=.9\columnwidth]{5_e2eMagLim_depSkyConditions.pdf}
	\fi
	\caption{%
		Long-exposure SR in H-band obtained for end-to-end simulations of the MICADO SCAO setup: with CLOSE (top), and with a manually optimized scalar integrator (bottom), for guide star magnitudes $M_R$=~11.5 -- 17.5 and seeing conditions $r_0$=~8.9 -- 21.5~cm.
	}
	\label{fig:5_statSeeing}
\end{figure}

Besides the adaptive capability of CLOSE shown in Sect.~\ref{sec:5_1_gainconvergence}, we investigate the steady-state AO performance achieved.
In order to perform this analysis, we generalize similar simulations as performed in Sect.~\ref{sec:5_1_gainconvergence} to a larger range of $r_0$ values (based on statistics provided by ESO within the frame of the development of ELT instruments) and guide star magnitudes (11.5 to 17.5).

Measured performances are shown on Fig.~\ref{fig:5_statSeeing} (top), with the long exposure H-band SR plotted against the star magnitude and computed for 5 different seeing conditions.
For all results the SR is averaged over 2-sec exposures, starting 3 sec after the AO loop is closed, and with initial modal gains $G_i = 0.5$ as previously.
Those simulations were also performed with a manually optimized scalar integrator ($G_i =$ constant, with one constant value per seeing and magnitude), with performance shown on Fig.~\ref{fig:5_statSeeing} (bottom), reasserting the improvement achieved using fine-tuned modal control demonstrated in previous research.
In particular, using modal control enables to (1) improve performance in poor seeing, even with bright guide stars, as nonlinearity is then a dominant member of the error budget. And (2), to push the performance envelope at the faint end, with a gain of typically +0.5 magnitudes for an identical objective.

While the simulation setup is not a perfect simulation of the MICADO system, Fig.~\ref{fig:5_statSeeing} (top) is indicative of the performance that could be achieved using CLOSE on such a system.
If we exclude the worst seeing conditions $r_0=8.9$~cm, a performance of 70\% SR or better is always achieved for bright stars, pushing to 89\% in excellent seeings; 40\% or better is achieved for $M_R$ up to 16.
SRs better than 40\% are also achieved for bright stars for $r_0=8.9$~cm; this potentially enables to salvage poor seeing nights for a number of scientific cases with milder correction requirements.

It is to be noted that CLOSE compares --for the metric used here-- equally to other, previously introduced modal compensation techniques for the PWFS \citep{Korkiakoski2008Improving, Deo2019Telescope, Chambouleyron2020Convolution}, but with the added value of automation and adaptability, without the offline computations, seeing estimations, dithering signals, or modifications to the optical setup required by those methods.

\subsection{Validation of the setpoint}
\label{sec:5_3_r_validation}

\begin{figure*}[t!]
	\centering
	\includegraphics[width=.99\textwidth]{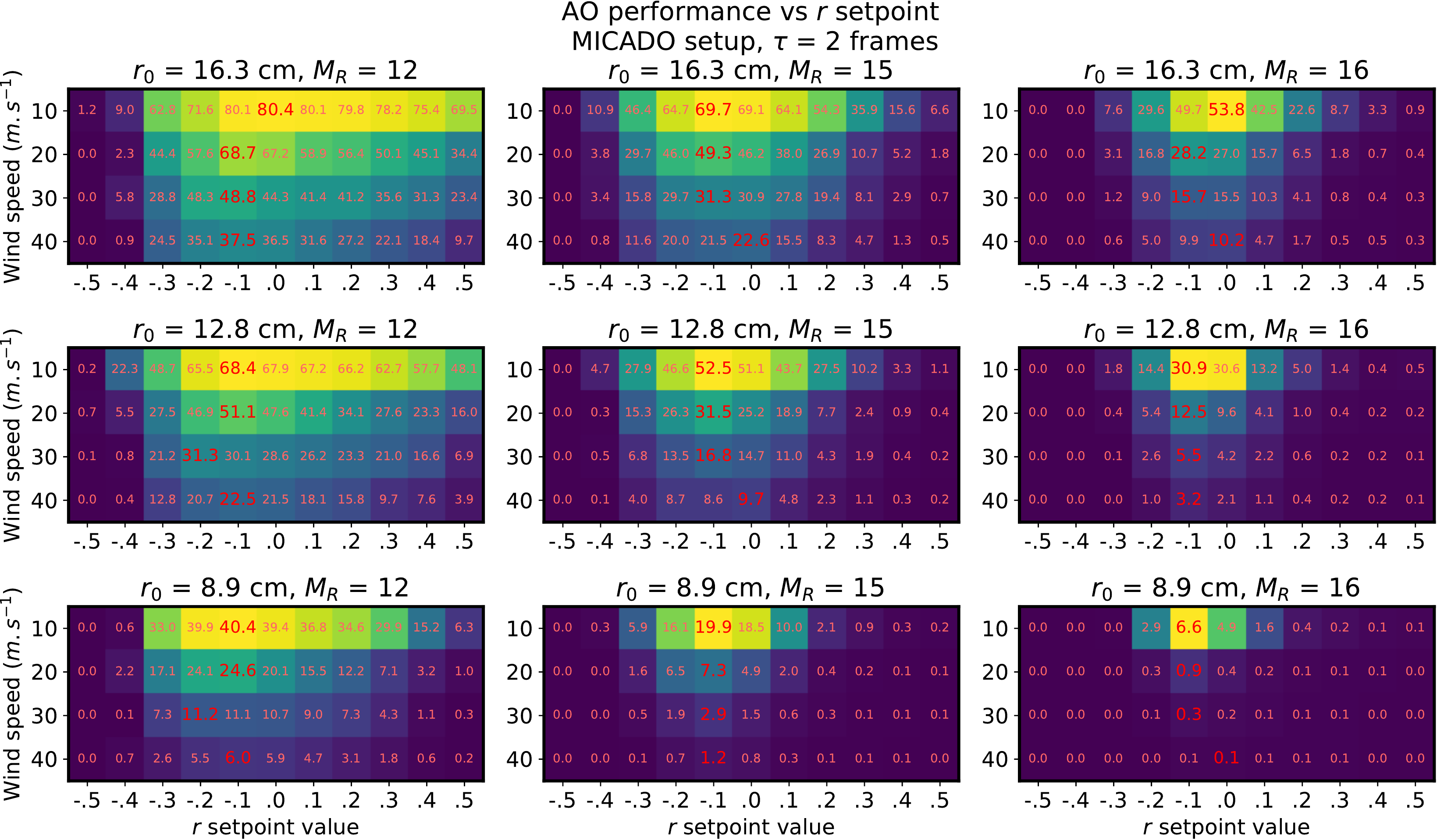}
	\caption{%
		Long exposure SR in H-band (color and text), for stationary simulations exploring different conditions regarding the seeing (outer rows), the guide star magnitude (outer columns), the wind speed (inner rows) and the CLOSE setpoint (inner columns), for the MICADO SCAO simulation.
		Color scales are local to each of the sub-plots and do not match each other.
	}
	\label{fig:5_rValues}
\end{figure*}

We have shown throughout Sect.~\ref{sec:3_closeTheory} and~\ref{sec:4_impl} that CLOSE is entirely configured with very few parameters; namely, $\dkc$, which ought to be chosen depending on the system latency; and $p$, $q^\pm$, $r$ being the choice of the operator.
While the temporal filtering introduced with $p$ and $q^\pm$ is easily apprehended and only impacts the performance during transitory regimes, the determination of $r$ is subject to more caution, as it determines the final performance.
Although semi-analytical computations indicated $r=0$ as a seemingly universal choice, those computations were the conclusion of a number of modeling hypotheses and approximations, as laid out in Sect.~\ref{sec:2_model}.

Our objective here is to validate whether $r=0$ stands as an appropriate choice in most situations.
We performed end-to-end simulations extending those presented in Sect.~\ref{sec:5_2_stationary}, now considering various $r$ setpoints from -0.5 to 0.5 and wind speeds from 10 to 40~m.s$^{-1}$.
The performances measured are shown on Fig.~\ref{fig:5_rValues}.
Some minor discrepancies can be found between Figs.~\ref{fig:5_statSeeing} and~\ref{fig:5_rValues}, explained by the different durations of simulated long exposures, and the mismatched random turbulence screens.
SRs for Fig.~\ref{fig:5_rValues} were computed on 600~msec exposures, following a 600~msec bootstrap for CLOSE and the AO. As compared to Fig.~\ref{fig:5_statSeeing} (2 + 2~sec), this was a necessary speedup given the large number of numerical simulations performed.
It should also be noted that LE SRs are generally determined to no better than 2-3\% of standard deviation over the distribution of turbulence screens.

Besides the expected variations of SR with seeing, guide star magnitude, and wind speed, we observe on Fig.~\ref{fig:5_rValues} that the setpoint yielding the maximum SR --$r_\mathrm{max}$ thereafter-- is almost always -0.1 or 0, except for two cases ($M_R=12$, $r_0=12.8$ and 8.9~cm and, 30 m.s$^{-1}$) where $r_\mathrm{max}=-0.2$.
When, for given $r_0$, $M_R$, and wind speed, we obtain $r_\mathrm{max} \neq 0$, we observe that the difference with the corresponding performance at $r=0$ generally amounts to 1-2\% only.
The slight bias towards $r_\mathrm{max} < 0$ may be partly explained by the reduced simulated time for CLOSE convergence, thus cases with lower $r$ would have increased the modal gains more effectively in the allocated 300 frames, due to the proportional effect introduced in Eq.~\ref{eq:4_GUpdate}.

The variations of the SR with $r$ are clearly determined from Fig.~\ref{fig:5_rValues}, as an asymmetric bell curve with a longer decrease on the $r > 0$ side.
With $r$ larger than the optimal value, the AO uses slower modal integrators with less turbulence rejection, which are more forgiving with regard to performance than $r$ being too small.
The latter cases introduce a buildup of noise that ultimately leads to diverging oscillations as $r$ gets too negative, hence a faster decrease in SR for simulations with $r$ smaller than the optimum.

A few outlying cases show a significant performance gap between the maximum SR and the SR achieved for $r=0$, up to a greatest difference of 6.3\%~($M_R=15$, $r_0=12.8~$~cm, 20 m.s$^{-1}$).
Such cases, where $r = 0$ induces a noticeably suboptimal performance, are all found for wind speeds of 20 m.s$^{-1}$ or more.
With the high latency of $\tau = 2$~frames simulated here, the higher wind speed induces a narrow acceptable range of gain $\alpha_iG_i$ for each mode to achieve near-optimal rejection, and thus a narrow range of $r$ values leading to this optimization with CLOSE.
As discussed in Sect.~\ref{sec:3_3_CLOSE_MV_comparisons}, while the CLOSE solution achieved for the modal gain is an empirical close match to the optimum rejection solution, we may be reaching the limits of this approximation for high latency, high wind speed cases.

To yet further investigate the usability of CLOSE with $r \approx 0$, we also performed similar simulations for different values of the system latency, using $\tau =$~0, 1 or 2 frames, and using AO setups other than the MICADO SCAO; namely two SCAOs on an 8~m telescope, respectively using a PWFS and a Shack-Hartmann (SH) WFS.
The configuration of these simulations and the obtained results are shown in Appendix~\ref{appdx:A_r_other}.
Beyond the results shown on Fig.~\ref{fig:5_rValues}, the extensive simulations exposed in Appendix~\ref{appdx:A_r_other} reinforce that the setpoint $r=0$ may be used as a baseline for a variety of AO systems, in a large number of observation situations.
The performance and steady-state behavior may eventually be fine-tuned by adjusting the $r$ setpoint upon empirical criteria if deemed necessary.

\subsection{Transients in observation conditions}
\label{sec:5_4_transients}

Beyond the performances of correlation-locking observed in stationary conditions exposed in Sects.~\ref{sec:5_2_stationary} and~\ref{sec:5_3_r_validation}, we propose here to analyze the dynamical capabilities of CLOSE in situations where the PWFS sensitivity or the illumination vary rapidly.
For that purpose, we designed two simulations with evolving conditions;
the first one emulates a seeing burst, with atmospheric conditions degrading dramatically during a short period; and the second one emulates a passing cirrus, with an equivalent drop of 3 magnitudes (-94\% of flux) of the guide star.

\begin{figure}
	\centering
	\ifreferee
			\includegraphics[width=.7\columnwidth]{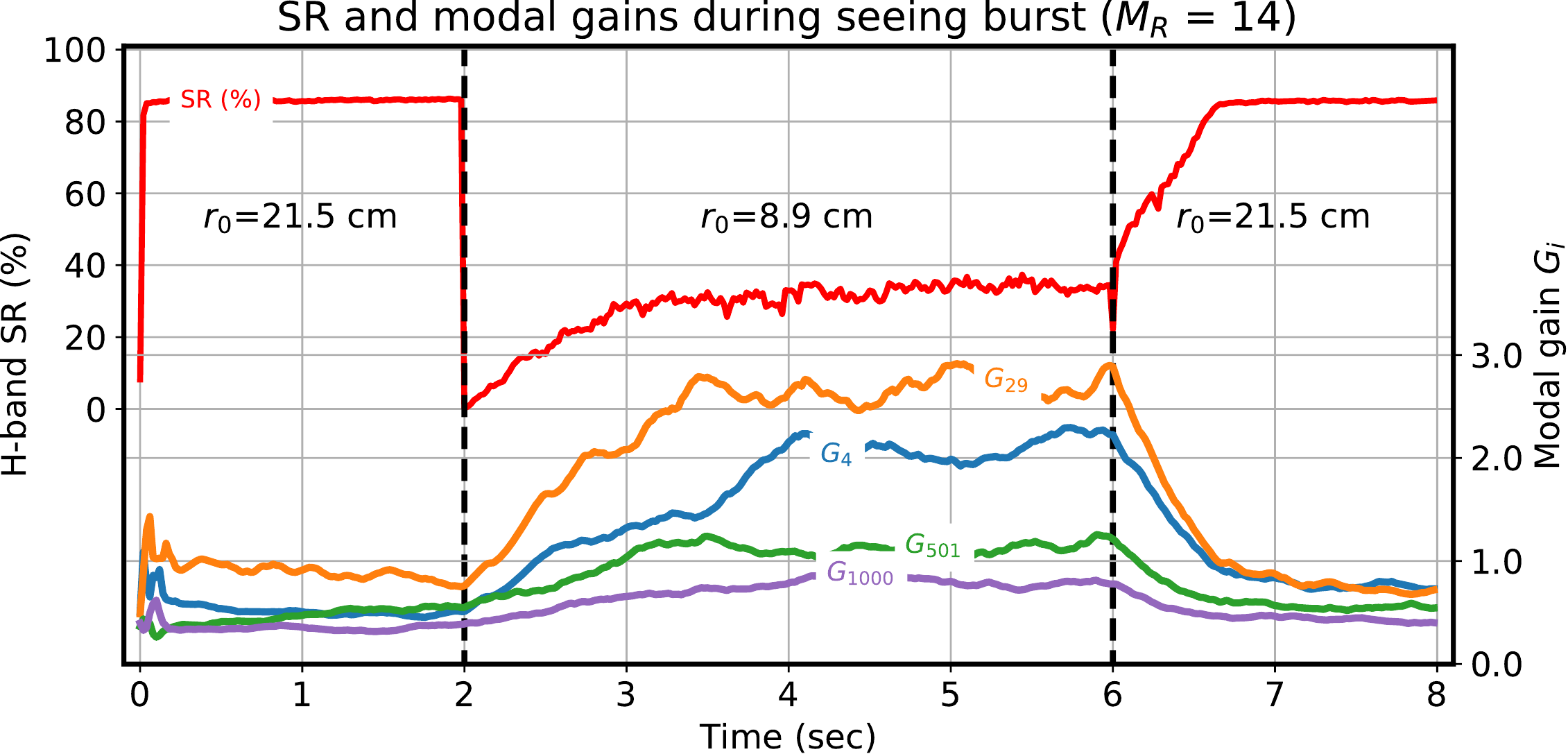}\\
			\includegraphics[width=.6\columnwidth]{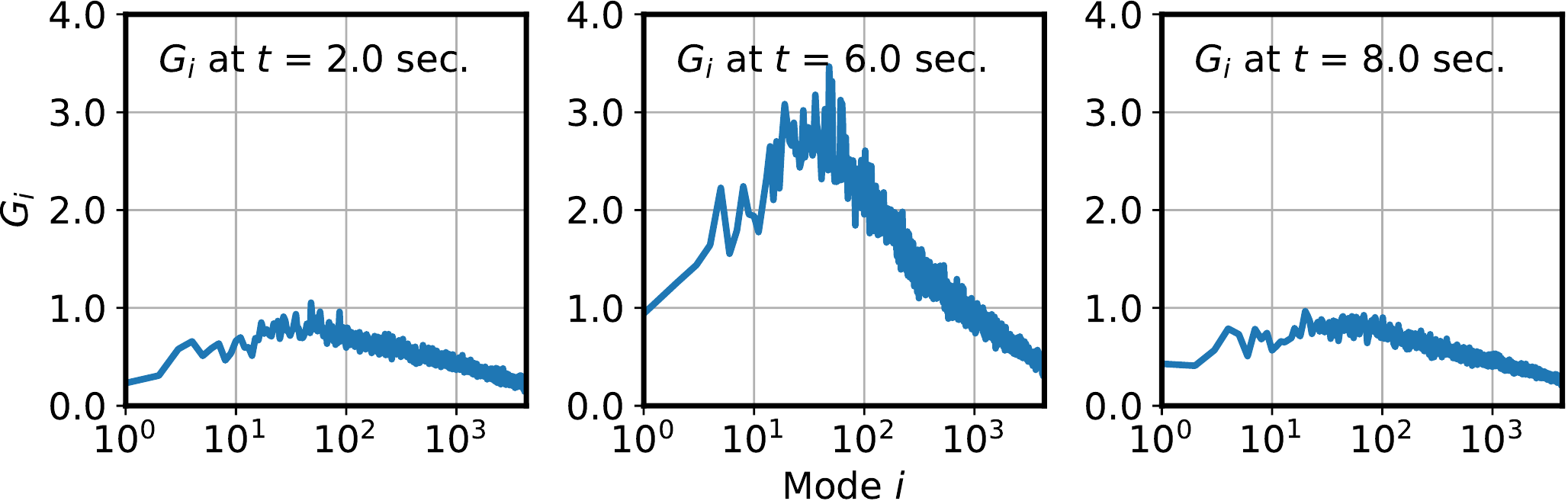}
	\else
			\includegraphics[width=\columnwidth]{5_4_r0transient.pdf}\\
			\includegraphics[width=.9\columnwidth]{5_4_r0transient_inserts.pdf}
	\fi
	\caption{%
		Top: time series of the H-band SR and modal gains $G_i$ for a few select modes, during the simulation of a burst of seeing; $r_0 = 21.5$~cm for $t < 2$~sec and $t > 6$~sec and $r_0 = 8.9$~cm for $2 < t < 6$~sec.
		Curves are smoothed using a 10~ms window.
		Bottom: snapshot of the modal gain vector $\vect{G}$ at time $t=$~2, 6 and 8 seconds.
	}
	\label{fig:5_seeingBurst}
\ifreferee % Merge the floats only in 2-col mode
	\end{figure}
	\begin{figure}
	\centering
\else
	\vspace*{1em}
\fi
	\ifreferee
		\includegraphics[width=.7\columnwidth]{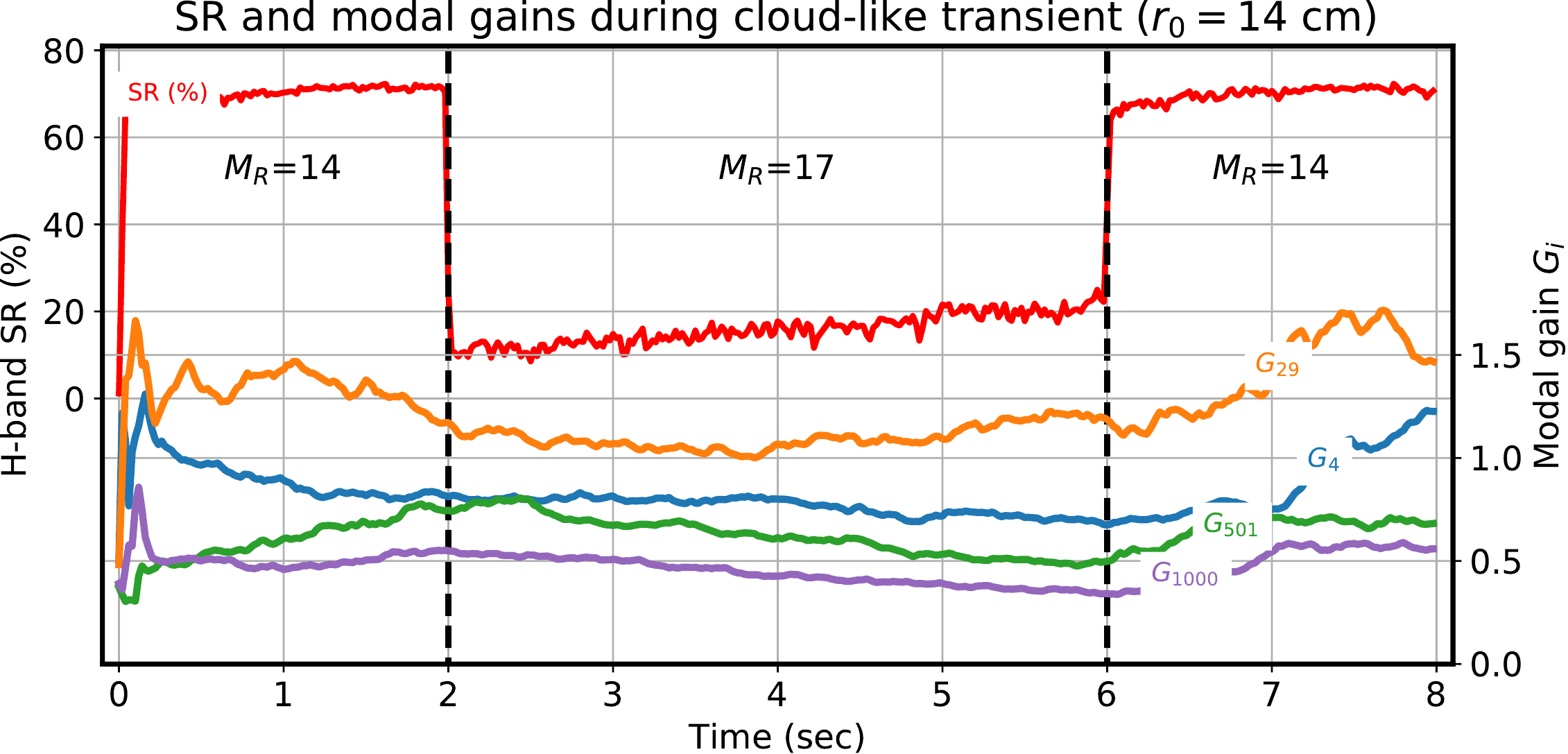}\\
		\includegraphics[width=.6\columnwidth]{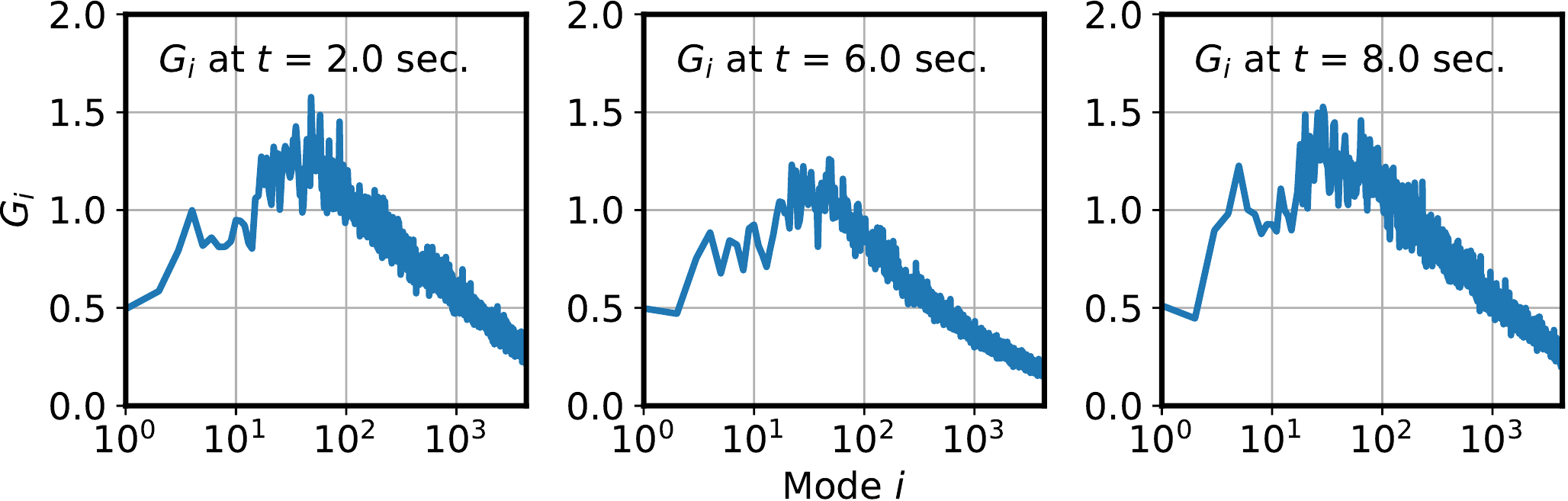}
	\else
		\includegraphics[width=\columnwidth]{5_4_cloudtransient.pdf}\\
		\includegraphics[width=.9\columnwidth]{5_4_cloudtransient_inserts.pdf}
	\fi
	\caption{%
		Similar to Fig.~\ref{fig:5_seeingBurst}. Time series and modal gain snapshots during a cloud-like transient, simulated as a guide star magnitude change: $M_R = 14$ for $t < 2$~sec and $t > 6$~sec and $M_R = 17$ for $2 < t < 6$~sec.
			}
	\label{fig:5_cloudBurst}
\end{figure}

Results for these simulations are shown on Figs.~\ref{fig:5_seeingBurst} and~\ref{fig:5_cloudBurst}.
We proceed identically for both transients, simulating 8~sec (4\,000 frames) of total runtime decomposed as: 2~sec of bootstrapping the AO loop from flat gains $G_i(t=0) = 0.5$; followed by 4~sec in degraded conditions; and finally 2 sec after reverting to the initial conditions.
Both Figs.~\ref{fig:5_seeingBurst} and \ref{fig:5_cloudBurst} show the time series of the H-band SR and of the modal gains $G_i$ for four select modes.
Lower panels in Figs.~\ref{fig:5_seeingBurst} and \ref{fig:5_cloudBurst} show a snapshot of the modal gain vector $\vect{G}$ over all 4\,301 DM modes at times $t=$~2, 6 and 8~sec, i.e., immediately before the changes in seeing or brightness, and at the end of the simulation.

\paragraph{Seeing burst}
For the seeing burst (Fig.~\ref{fig:5_seeingBurst}), the turbulence screen is scaled up between $t=2$ and 6~sec to simulate a $r_0$ change from 21.5~cm to 8.9~cm, while keeping a guide star of brightness $M_R=14$.
The sudden changes in $r_0$ at $t=2$~and 6~sec induce discontinuities in the wavefront --as we perform an instantaneous scaling across the entire aperture--, inducing very short drops to 0\% SR (a few frames).
It can be seen that at $t=2$~sec, the modal gains optimized for $r_0=21.5~$cm do not permit to reach optimal performance immediately now that $r_0=8.9~$cm;
for $2<t<3$~sec, CLOSE drives the $G_i$ up by a factor 2 to 4, to accommodate the OG-induced sensitivity reduction, significantly more important at $r_0=8.9~$cm.
This second long convergence to the new optimal state allows a SR improvement up to the 30-35\% consistently measured for $3<t<6~$sec, which is the expected performance in such stationary conditions ($r_0=8.9$~cm, $M_R=14$; as seen in Fig.~\ref{fig:5_statSeeing}).

Similarly, the second transition back to $r_0 = 21.5~$cm at $t=6$~sec shows an immediate SR performance about 45\% below the expectation, indicating the system is out of tune, as the $G_i$ are now beyond reasonable values for $r_0=21.5~$cm.
Yet, the modal gains are automatically decreased, and the SR converges to ~85\%, within 0.5~sec.
This is consistent with the magnitude of the time constants involved, as we use a learning factor $q^\pm=10^{-3}$ at a framerate of 500~Hz.

\paragraph{Photometric variations}
The ``cirrus simulation'' (Fig.~\ref{fig:5_cloudBurst}) is run using the median seeing $r_0 =~14.5$~cm, with a change of guide star brightness from the initial $M_R=14$ down to $M_R=17$ during the perturbation interval $2<t<6$~sec.
While such a flux attenuation qualitatively appears as a major change, the impact on the gain is rather mild when compared to the previous case.
The reaction of CLOSE that can be observed on Fig.~\ref{fig:5_cloudBurst} is that an appropriate overall reduction in modal gains is progressively introduced during the transient, and then reverted after $t=6$~sec.
Such behavior is typically what is expected from modal gain optimization with S/N; looking at the SR temporal evolution, we can see that the progressive optimization of the transfer functions for $M_R=17$ allows for an improvement from 10 to 20\% of SR within the 4 seconds spent at this brightness level.

It is worth noting that while the modal gain curve reached at $t=6$~sec shows only a modest reduction as compared to $t=2$~sec, this reduction is in fact the product of two counterbalancing effects: first, a reduction of $\alpha_i G_i$ for improving the noise rejection; but coupled to an increase in the sensitivity compensation $\alpha_i^{-1}$, as the mean residual wavefront is notably increased, thus inducing more OG sensitivity reduction in the PWFS response.
These two effects, of course, are entirely implicit:
as the flux drops by 3 magnitudes, the corrective capability of the AO immediately decreases, the residual wavefront increases, and thus the sensitivity $\alpha_i$ decreases as well.
Factored in the transfer function, the gain $\alpha_iG_i$ is decreased immediately and compensates for the sudden shift in the S/N. Afterwards, CLOSE accounts for the fine-tuning of the $G_i$ to more precisely set the rejection at the new sensitivity level.
The opposite happens at $t=6$~sec: the sharpening of the correction improves the sensitivity, and immediately the dynamical gain $\alpha_iG_i$ sees an increase, and provides more turbulence rejection, now with less noise to amplify.
This effects explains why, as compared to the ``seeing burst'', (1) the fluctuations in $G_i$ are significantly milder; and (2) why the return-to-steady-state time is much shorter.

Finally, as with the seeing burst, we note that the performance in the final state $t>6$~sec is identical to the beginning $t<2$~sec, pointing that no diverging signals were accumulated during the transitions and that the AO is capable of reverting cleanly after such transients.

\paragraph{Synthesis}
From the two experiments performed with sudden changes in observing conditions,
we may conclude that CLOSE efficiently and automatically accommodates sudden, significant changes in S/N, whether from the seeing or the guide star brightness.
This robustness to transients is observed both for improving or deteriorating conditions.
This efficient convergence towards nominal performance, when the seeing improves in particular, underlines the robustness of the proposed AO control scheme;
when the seeing suddenly improves, the AO is configured with gains significantly too high for the newly set turbulent conditions, yet the control scheme allows to re-optimize the gains without the loop diverging.

%  ____  _                        _                 
% |  _ \(_)___  ___ _   _ ___ ___(_) ___  _ __  ___ 
% | | | | / __|/ __| | | / __/ __| |/ _ \| '_ \/ __|
% | |_| | \__ \ (__| |_| \__ \__ \ | (_) | | | \__ \
% |____/|_|___/\___|\__,_|___/___/_|\___/|_| |_|___/
\section{Discussion}
\label{sec:6_discu}

We propose in this Section a few additional comments on the CLOSE algorithm, its fundamental behavior, and some of its foreseen limitations at this stage of development.

\subsection{In general: an interaction matrix transform}
\label{sec:6_1}

In the most general sense, as developed in Sects.~\ref{sec:2_model} and~\ref{sec:3_closeTheory}, CLOSE is an automatic minimum variance controller for servo-loops which input is built of a low-frequency rumble and a white noise floor (Eq.~\ref{eq:2_3_turbSpectrum}, Fig.~\ref{fig:3_spectralOutput}).
When such conditions are met, as is the case with astronomical adaptive optics, and when independent modal loops can be decoupled, CLOSE is an excellent tool to optimize the system regardless of miscalibrated sensitivity factors.

As such, CLOSE is adequate for any system which admits a description as in Fig.~\ref{fig:2_globalSchem}, where an eigenbasis can be found that properly describes the nonlinearity of the system, i.e., diagonalizes (well enough) the modal confusion matrix $\textbf{A}$.
This is the case for the PWFS, as has been proven experimentally \citep{Deo2018Modal} and analytically \citep{Fauvarque2019Kernel, Chambouleyron2020Convolution}; and this has been extended to the entire class of Fourier-based wavefront sensors.

Furthermore, it has been demonstrated that a number of configuration changes of PWFSs are very well described by variations of the modal matrix $\mathbf{A}$, while the interaction matrix $\mathbf{dWFS}$ can be kept to the unresolved, bright, on-axis, single star one acquired in the lab (or simulated).
Such changes include online changes to the modulation radius, or tracking on double stars of various separations and orientations relative to the pyramidal prism edges, as well as on other extended objects~(Titan, uncompensated atmospheric dispersion beacon; \citealp{Vidal2019Analysis}).
For binary stars, it has been in particular shown that using CLOSE restores the performance of the servo-loop to near single-star levels, without the additional burden of biasing the loop setpoint to drive one binary component to the PWFS apex, which is the usual --and nonetheless efficient-- alternative.

All of these transformations, similarly to optical gain, can be approximated by Fourier filters altering the description of the sensor, and thus are adequately tracked by CLOSE.
Ideally, a Fourier basis would have to be used to properly diagonalize $\mathbf{A}$;
our experience and numerical analyses demonstrate that a frequency-ordered KL basis is a sufficient approximation.
We have previously studied~\citep{Deo2018Modal} --for the case of optical gain only-- that the scaling applied by $\mathbf{A}$ on Fourier modes of identical norm is isotropic within a few percent, thus proving KL modes are adequate to such an operation.

\subsection{Further required steps}
\label{sec:6_2}

Finally, we would like to point to a few elements that are currently beyond the scope of what CLOSE is capable of achieving.
One of the most prominent issues with high order PWFSs is the proper subtraction of non-common path aberrations (NCPAs) as the servo-loop setpoint, which issue is tightly coupled with the estimation of modal OG compensation coefficients.
In the state presented in this paper, CLOSE does not enable an independent estimation of the transfer function gain $\alpha_iG_i$ and the OG compensation coefficient $\alpha_i^{-1}$ --as has been observed in the ``cloud transient'' analyzed in Sect.~\ref{sec:5_4_transients}.
Obtaining $\alpha_i^{-1}$ is necessary for a proper OG-adjusted compensation of the setpoint.
Other techniques, which include injecting probe signals on the system DM, allow an explicit retrieval of $\alpha_i^{-1}$~\citep{Esposito2020NCPA}.
While CLOSE does not include this feature, it is very compatible with such techniques, while being much more reactive to short term transients. A joint implementation of several modal gain estimation techniques would be extremely beneficial for an all-in-one solution for PWFSs control, including minimum variance laws, proper NCPA compensation, shift and misalignment tracking~\citep{Heritier2018Analysis}, etc.

One additional comment should be made on the potential limitations of the modeling laid in Sect.~\ref{sec:2_model}.
The proper variance minimization obtained with CLOSE relies on the clear separation of two frequency components from the turbulence and the noise in the resulting measurement PSD: $|\hat{m}_i|^2(f)$.
The requirements of this paper go further, requiring that the modal noise is white and that the turbulence spectrum obeys Eq.~\ref{eq:2_3_turbSpectrum}.
We have however proven with the Shack-Hartmann results shown in Appendix~\ref{appdx:A_r_other} that the decay exponent $-17/3$ of Eq.~\ref{eq:2_3_turbSpectrum} may allow some tolerance -as it is then actually $-11/3$ due to additional subaperture aliasing.
This, as well as using different von Kármán outer scales $L_0$, is very likely to affect the quantitative conclusions of the semi-analytical derivations led in Sect.~\ref{sec:3_3_CLOSE_MV_comparisons}, however we are confident that it will not compromise the monotonicity of the relationship between the $\dkc$-correlation and the loop gain $\alpha_i G_i$.
Thus, a simple re-assessment of the loop setpoint $r$ to try and achieve quasi-optimality is likely to be sufficient for such situations.

Telescope vibrations are a less trivial modification of the input turbulent spectrum. Naturally, CLOSE cannot achieve any better vibration rejection than other non-predictive integrators. As for other spectrum modifications, a re-assessment of $r$ could help in achieving near optimality, as good as this may be in the presence of narrow vibration peaks.
Low frequency (e.g., <10~Hz) vibrations would be implicitly registered as a degradation in seeing, and would require to bias the setpoint towards $r>0$ to prevent the corresponding gain increase. Conversely, vibrations between $\fc / 2$ and $3\fc / 2$ would be registered as noise energy, requiring to bias the setpoint towards $r<0$.

We foresee vibrations at or near $\fc$ to cause specific issues.
All of our simulations have shown that CLOSE, seeking its near-MV solutions, tends to push the loop gain very close to its critical value when the S/N is good enough.
This could lead to transients with very high vibration amplification, e.g., as the vibration restarts after a pause during which $\alpha_i G_i$ was boosted near $g_\mathrm{crit}$.
Such situations will require the implementation of specific fail-safes to avoid the sudden divergence of the AO loop.
More generally, CLOSE in the presented version does not currently cope well with spurious divergences of the AO loop: the loss of WFS feedback associated with DM divergence (or clipping at maximum value) will induce CLOSE (if $r\leq0$) to keep increasing gains towards infinity, even though the loop has already diverged.

It will now be necessary to experiment with various modifications of the input spectra (correlated noise, vibrations, non von Kármán spectra) and the integral command law (leaky integrators, linear predictive methods), as to investigate to what extent and with what versatility we may obtain satisfactory results with the version of CLOSE presented in this paper, and which further improvements will prove necessary.

%  ____                 _           _             
% / ___|___  _ __   ___| |_   _ ___(_) ___  _ __  
% | |  / _ \| '_ \ / __| | | | / __| |/ _ \| '_ \ 
% | |_| (_) | | | | (__| | |_| \__ \ | (_) | | | |
% \____\___/|_| |_|\___|_|\__,_|___/_|\___/|_| |_|
\section*{Conclusion}
\label{sec:conclusion}

We have presented a self-regulating adaptive filtering method that enables the tracking, either in hard or soft real-time, of the modal gains of the integrator controller in an AO system.
This technique allows to keep the system always close to peak performance against variable conditions (seeing, wind speed, S/N) and the variable sensitivity of the WFS.
Being entirely automatic and independent of WFS sensitivity fluctuations, the proposed scheme is particularly well suited for high-order PWFS designs on ELTs, which are prone to strong optical gain effects.
Yet, this OG effect is well decoupled and stable against seeing statistics using a KL control basis of the DM, as we use, thus enabling a separable modal analysis throughout.

Our method has been called CLOSE, for Correlation-Locking Optimization SchemE.
It leverages the monotonic relationship between the hidden, actual transfer function gain and the loop resonance peak, which amplitude is indirectly assessed through the temporal auto-correlation of the modal phase residuals computed at a single given time-shift $\dkc$.
The appropriate counter-reaction is then applied on the modal gain in order to \emph{lock} the previously mentioned $\dkc$-correlation value to a chosen setpoint.
Using semi-analytical computations, we have demonstrated that for given conditions, there is an unequivocal relation between the transfer function gain and the value of the $\dkc$-correlation; and that correlation-locking easily permits to almost achieve a modal minimum variance criterion regardless of the modal S/N.

We have used extensive end-to-end AO simulations to show the versatility of CLOSE. Using the unique setpoint $r=0$, we have shown automatic convergence of the modal gains in an wide variety of simulated conditions, using different wind speeds, guide star magnitudes, and seeing parameters.
We have also demonstrated the automatic adaptability of the scheme to abrupt changes in observation conditions, hopping between different sets of optimal control gains in a matter of seconds.
These simulations use the design parameters of the PWFS of the MICADO SCAO system, and, compared to recent design studies for this system~\citep{Vidal2019Analysis}, achieve nominal performance accounting both for optical gain modal compensation and modal transfer function optimization. This performance is achieved while entirely alleviating from extensive situation-dependent optimizations, database look-ups, or manual tune-ups, thus offering CLOSE as an all-in-one baseline strategy for AO modal control.

While CLOSE elegantly paves way for solving variance minimization issues with nonlinear sensors, we raised in Sect.~\ref{sec:6_2} some warnings as to what the scheme does not provide in the version described in this paper: neither an explicit estimation of the sensitivity reduction coefficient, nor a fully fledged solution to the NCPA subtraction issue for sensors of varying sensitivity.

Work is currently underway to refine and expand the work described in this paper to make it usable for on-sky operations.
Importantly, we shall design and deploy the necessary fail-safes in case CLOSE suffers from a loss of reliable feedback from the modal outputs, or if the AO falls into non-linear measurement pits.
In such cases, the modal gain update equations tend to increase the gains exponentially.
These situations must be automatically identified and unambiguously distinguished from transient sensitivity reductions.
Should its capabilities, convenience, and versatility be confirmed, we envision CLOSE as a core control technique for future AO systems, embedded within a larger framework of leveraging real-time telemetry for AO control.

\begin{acknowledgements}
This research is performed in the frame of the development of MICADO, first light instrument of the ELT (ESO), with the support of ESO, INSU/CNRS and Observatoire de Paris.
V.~Deo is supported by NASA grant 80NSSC19K0336.
The authors thank the COMPASS development and support team for their continued support.
\end{acknowledgements}

% References
% Doctorat abbreviated is better yet not up to date.
\bibliography{doctorat}
\bibliographystyle{aa}

% Appendices - stuff from SPIE ?
\appendix
\renewcommand{\theequation}{\thesection.\arabic{equation}}
\numberwithin{equation}{section}
\renewcommand{\thefigure}{\thesection.\arabic{equation}}
\numberwithin{figure}{section}

\section{Expanding on AO transfer functions}
\label{appdx:C_CLOSE_TransferFunc}

We detail here, as a complement to Sect.~\ref{sec:2_3_transferFunctions}, the transfer functions of the AO loop and highlight the differences between a continuous-time description, as is necessary for a real, physical system, and a discrete-time description, adequate for AO simulators such as COMPASS.
The material reviewed in Sect.~\ref{appdx:C_1} is well known (e.g., \citealp{Madec1999Wavefront,Kulcsar2006Optimal}), yet seldom presented, and we deem it necessary for a proper mathematical grounding of Sects.~\ref{sec:2_3_transferFunctions} and~\ref{sec:3_closeTheory}.
Ultimately, this development lets us derive the values for $f_\mathrm{crit}$ and $g_\mathrm{crit}$ provided in Table~\ref{tab:3_summary}.

\subsection{Continuous- and discrete-time descriptions}
\label{appdx:C_1}

Both the real, continuous-time system (CTS) and the simulated, discrete-time system (DTS) are clocked at the sampling period $T$, which is also the integration time of the WFS detector.
%In the DTS, the turbulence is not averaged over $T$, but sampled at a single time; similarly the action of the DM is not held during a duration $T$, ``but combined as a sample to another sample'' \vdeo{What does that mean?}.
In the DTS, the turbulence is not averaged over $ T $, but sampled at a single instant; likewise, the notion of the duration of action of the deformable mirror simply does not exist.
Both systems, however, have in common the core of the control: they share the same numerical algorithms implemented in the real-time computer, that are in both cases numerical and in discrete time steps.

We introduce several transfer functions hereafter. The WFS performs an integration of the signal during $T$.
In Fourier formalism, its transfer function can be written as
\begin{equation}
\label{eq:C_hwfs}
h_\mathrm{wfs}(f) = \exp\left(-j\pi f T\right) \frac{\sin\left(\pi f T\right)}{\pi f T}
\end{equation}
which is the Fourier transform of a rectangular window extending from $t=0$ to $t=T$.
We may see Eq.~\ref{eq:C_hwfs} as a smoothing by a zero-centered box: $\mathrm{sinc}(\pi f T)$, combined with a half-frame delay: $\exp\left(-j\pi f T\right)$, thus ensuring $h_\mathrm{wfs}$ describes a causal system.

The WFS measurement is processed by the reconstructor matrix to be expressed in a modal space $m[k]$ --we drop here the mode index $i$ for clarity--, and is integrated with a gain $g$ into the modal command $v[k]$, using the classical discrete integrator:
\begin{equation}
\label{eq:C_digInteg}
v[k] = v[k-1] + g\cdot m[k],
\end{equation} 
where $k$ is the loop iteration number.
%\vdeo{Can you clarify / shorten the following ? The variable $k$ should not be confused with an index representing time: it only designates the numerical iterations in the real-time computer, it only establishes a cause and effect relationship between the variables $v[.]$ and $m[.]$. For that reason, this equation does not tell anything about the time aspects between $v[k]$ and $m[k]$. We will introduce later the delay required to shift the samples $v[k]$ and $m[k]$ on the time axis.}
The direct translation of Eq.~\ref{eq:C_digInteg} in the frequency space gives the expression of the transfer function of the discrete time integrator:
\begin{equation}
\label{eq:C_numericalIntegrator}
g \cdot h_\mathrm{digInt}(f) = \dfrac{\hat{v}(f)}{\hat{m}(f)} = \dfrac{g}{1 - \exp(-2j \pi f T)}\,.
\end{equation}
It should be noted that $h_\mathrm{digInt}(f)$ is nonphysical, first because it shows no computational time delay; and second because it is periodically undefined for every multiple of the sampling frequency $f_S=1/T$ : it is only appropriate for representing a discrete time system.

Now, in an real CTS, the RTC computation result $v[k]$ is sent to a digital-to-analog converter (DAC), which holds the (modal) command on the DM during the period $T$. The DAC --a zero-order hold-- has the same transfer function as the WFS:
\begin{equation}
\label{eq:C_Hdacs}
h_\mathrm{dac}(f) = \dfrac{1 - \exp\left(-2j\pi f T\right)}{2j \pi f T},
\end{equation}
also including a half-frame delay.
With the association $h_\mathrm{wfs}\cdot h_\mathrm{dac}$, a \emph{full} frame delay thus naturally appears.

Eq.~\ref{eq:C_Hdacs} joins the digital world back into the physical reality of the CTS, by transforming the Dirac comb of numerical samples into an analog signal.
Noticeably, the association of the ``non-physical'' numerical integrator with the DAC transfer function form the following product:
\begin{equation}
h_\mathrm{digInt}(f).h_\mathrm{dac}(f) = \frac{1}{2j \pi f T},
\end{equation}
i.e., the proper transfer function of a continuous-time integrator.

The final point of discussion is that of the time delay $\tau$.
We have defined the loop delay as the \emph{additional} amount of time, on top of the causal process.
For a CTS, this is the time spent between the end of the WFS integration, and the beginning of the application of the command on the DM.
In such case, the transfer function is:
\begin{equation}
h_\mathrm{delay}(f;\,\tau) = \exp(-2j \pi f \tau).
\end{equation}
When $\tau=0$, the DM command $v[k]$ exactly applies during the acquisition of $m[k+1]$.
This one-frame shift is built in the equations thanks to the two half-frames from the WFS and DAC.
All the elements are introduced for the transfer function between the input turbulence $\phi^i_\mathrm{Atm}$ and the modal measurements $m_i[k]$ in the case of a CTS:
\begin{equation}
\label{eq:C_hcorrcts}
h_\mathrm{corr,CTS}(f;\,g) = \dfrac{1}{1+g\cdot h_\mathrm{wfs}(f)\cdot h_\mathrm{digInt}(f)\cdot h_\mathrm{dac}(f)\cdot h_\mathrm{delay}(f;\,\tau)}.
\end{equation}

For the DTS, there is neither WFS nor DAC windowing, but we need to introduce a seemingly artificial one-frame delay $\exp(-2 j \pi f T)$ to account for the fact that a command computed at a given iteration can, at best, only serve for the next one:
\begin{equation}
\label{eq:C_hcorrdts}
h_\mathrm{corr,DTS}(f;\,g) = \dfrac{1}{1 + g\cdot h_\mathrm{digInt}(f)\cdot h_\mathrm{delay}(f;\,\tau)\cdot \exp(-2j \pi f T)}.
\end{equation}
Expanding, we obtain:
\begin{equation}
h_\mathrm{corr,DTS}(f;\,g) = \dfrac{1}{1 + g\cdot \dfrac{\exp(-2j \pi f (T+\tau))}{1-\exp(-2j \pi f T)}},
\end{equation}
i.e., the shorthand $h(f; g)$ of Eq.~\ref{eq:2_hi}.
%\vdeo{Rico: peux-tu conclure par un bref commentaire sur pourquoi on peut presque dire que c'est tout pareil et qu'on peux utiliser la DTS dans le main texte sans se poser de questions ?}

With the closed-loop transfer functions with a CTS and a DTS representation of the AO, let us now study the \emph{critical} behavior of the system, as the gain increases to approach divergence.

\subsection{Critical point -- analog case}
\label{appdx:C_2}

The critical frequency and gain (Sect.~\ref{sec:3_1_rationale}) are determined by finding the couple of $g_\mathrm{crit}$ and $f_\mathrm{crit}$ that zero the denominator of Eq.~\ref{eq:C_hcorrcts}.
The equation to be solved for $g$ and $f$ is:
\begin{equation}
g \dfrac{1}{2j \pi f T}  \exp\left(-j\pi f T\right)\dfrac{\sin\left(\pi f T\right)}{\pi f T} \exp(-2j \pi f \tau) = -1.
\end{equation}
Equating for phase and modulus:
\begin{equation}
g \dfrac{\sin(\pi f T)}{2 \pi^2 f^2 T^2} = 1\text{,\ \ and\ \ }
-\frac{\pi}{2} - \pi f T - 2 \pi f \tau = - \pi,
\end{equation}
which reduce to
\begin{align}
f_\mathrm{crit}  & = \frac{1}{2T+4\tau} \\
\label{eq:C_gcritanalog}
g_\mathrm{crit} & = \frac{2 \pi^2 f_\mathrm{crit}^2 T^2}{\sin(\pi f_\mathrm{crit} T)}    
\end{align}

\subsection{Critical point -- digital case}
\label{appdx:C_3}

Same as for Sect.~\ref{appdx:C_2}, the equation to be solved is
\begin{equation}
g \dfrac{\exp(-2j \pi f (T+\tau))}{1-\exp(-2j \pi f T)} = -1.
\end{equation}
Leveraging
\begin{align*}
1-\exp(-2j \pi f T)  & = 2 j \exp(-j \pi f T)\sin(\pi f T)         \\
& = 2 \exp\left(j \dfrac{\pi}{2}\right) \exp(-j \pi f T)\sin(\pi f T),
\end{align*}
we obtain the phase and modulus equations:
\begin{equation}
g = 2 \sin(\pi f T)\text{,\ \ and\ \ }
-2 \pi f \tau -2 \pi f T = -\pi  + \pi/2 - \pi f T,
\end{equation}
and ultimately,
\begin{align}
f_\mathrm{crit}  & = \frac{1}{2T+4\tau}          \\
\label{eq:C_gcritdiscrete}
g_\mathrm{crit}  & = 2 \sin(\pi f_\mathrm{crit} T), 
\end{align}
which is the general formula given in Table~\ref{tab:3_summary}.
It should finally be noted that for a CTS with latency $\tau \geq 1$, we have $f_\mathrm{crit}T < 1/6 << 1$, and therefore Eqs.~\ref{eq:C_gcritanalog} and ~\ref{eq:C_gcritdiscrete} both reduce to the same value $g_\mathrm{crit}  \approx 2 \pi f_\mathrm{crit} T$ in a first order approximation.
The general discussions and derivations presented in Sect.~\ref{sec:3_closeTheory} are in that case also applicable to a continuous-time description of the system.

And, as a final note, \citet{Kulcsar2006Optimal} lays out the necessary hypotheses and follows to prove that a discrete-time treatment of the minimization criterion in Eq.~\ref{eq:2_3_PSDVariance} is equivalent to a continous-time treatment.

\section{Semi-analytical CLOSE solutions for latencies of 0 and 1 frames}
\label{appdx:B_CLOSE_MV}

\begin{figure}[t]
	\centering
	\ifreferee
	\includegraphics[width=.65\columnwidth]{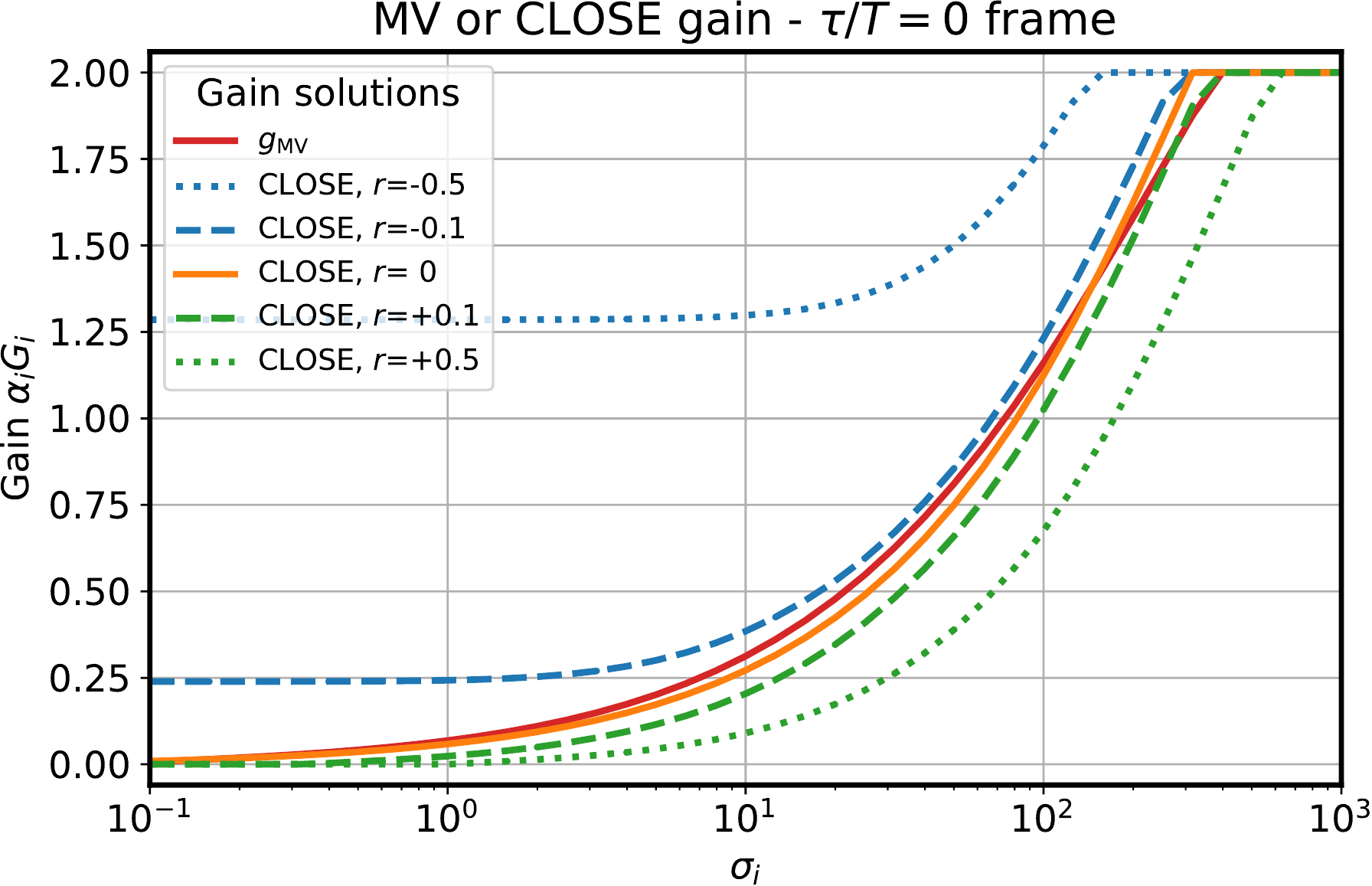}
	\else
	\includegraphics[width=.98\columnwidth]{3_CLOSE_MV_lat0.pdf}
	\fi
	\caption{%
		Minimum variance gain $g_\mathrm{MV}$ and solutions found using CLOSE with five different setpoint values, depending on the sensitivity corrected S/N $\sigma_i$.
		These computations are performed using the modal input spectrum shown on Fig.~\ref{fig:3_spectralOutput}.
	}
	\label{fig:B_CLOSE_MV_comp_lat0}
\end{figure}

\begin{figure}[t]
	\centering
	\ifreferee
	\includegraphics[width=.65\columnwidth]{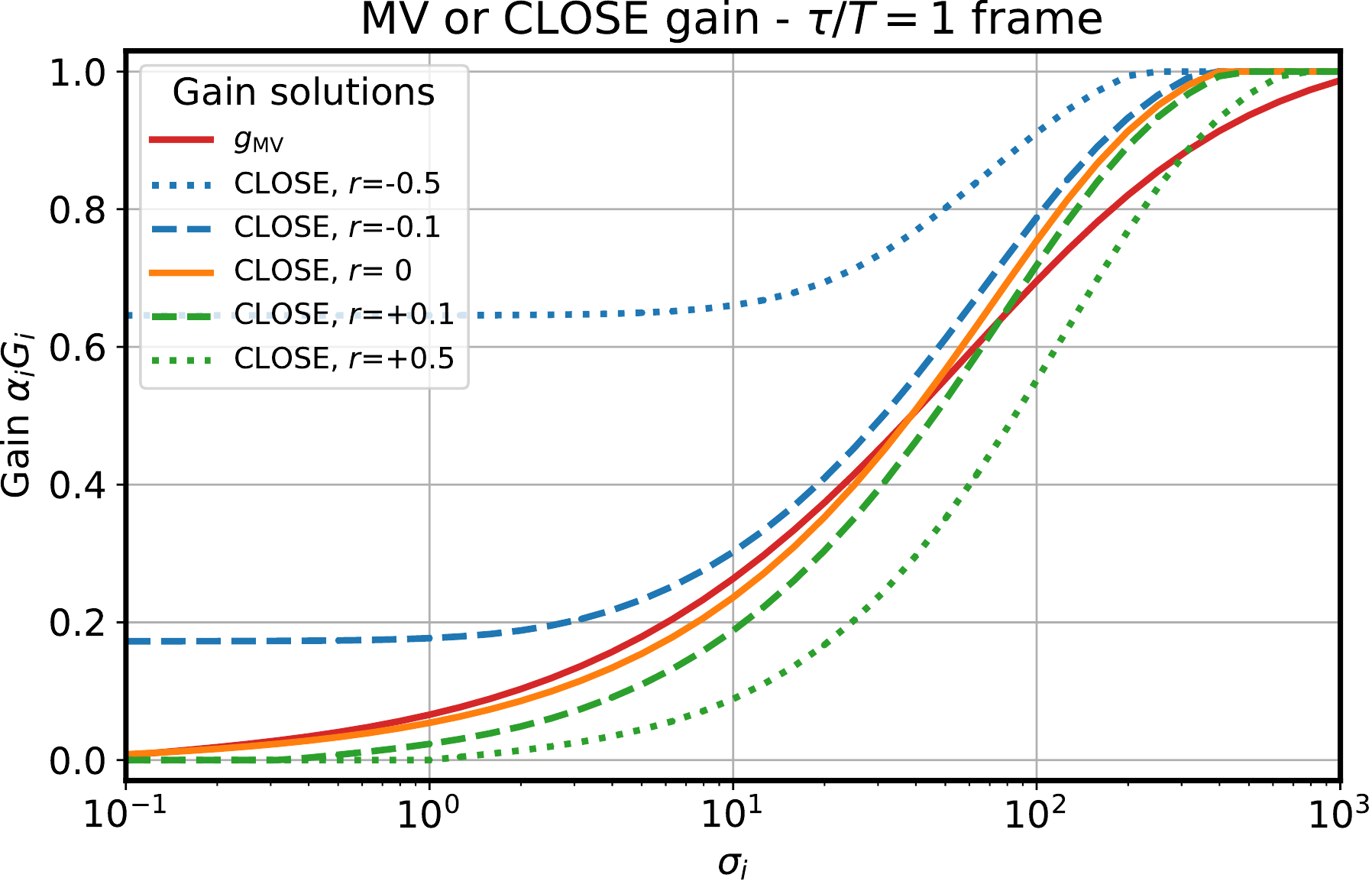}
	\else
	\includegraphics[width=.98\columnwidth]{3_CLOSE_MV_lat1.pdf}
	\fi
	\caption{%
		Minimum variance gain $g_\mathrm{MV}$ and solutions found using CLOSE with five different setpoint values, depending on the sensitivity corrected S/N $\sigma_i$.
		These computations are performed using the modal input spectrum shown on Fig.~\ref{fig:3_spectralOutput}.
	}
	\label{fig:B_CLOSE_MV_comp_lat1}
\end{figure}

We propose in this appendix some additional data expanding the discussion led in Sect.~\ref{sec:3_3_CLOSE_MV_comparisons}, exploring the close match obtained --for all sensitivity-corrected S/N $\sigma_i$-- between the minimum variance gain $g_\mathrm{MV}$ and the steady-state gain yielded by CLOSE for a setpoint $r=0$.
While in Fig.~\ref{fig:3_CLOSE_MV_comp} this comparison was restricted to a latency $\tau = 2T$, we show on Figs.~\ref{fig:B_CLOSE_MV_comp_lat0} and~\ref{fig:B_CLOSE_MV_comp_lat1} the same analysis for latencies $\tau = 0$ and $\tau = T$.
All parameters are otherwise similar, in particular the sample atmospheric spectrum (Eq.~\ref{eq:2_3_turbSpectrum}) with a cutoff of 1~Hz.
Of course, the time-difference $\dkc$ at which the autocorrelation is locked to the setpoint $r$ is taken per Table~\ref{tab:3_summary}.

For all three latencies $\tau=0$, $T$ and $2T$, we observe the same remarkable match between the minimum variance solution and CLOSE for a null setpoint.
This has also been simulated for integer latencies up to $\tau = 5T$, although such cases are not necessarily relevant to a real AO system and are not reported here.
While Fig.~\ref{fig:B_CLOSE_MV_comp_lat0} reports a very satisfactory correspondence, it should be mentioned that the analysis proposed reaches the limit of the discrete time approximation of real AO systems, as is explicited throughout Appendix~\ref{appdx:C_CLOSE_TransferFunc}.
With $f_\mathrm{crit} = f_S / 2$, the case $\tau=0$ is therefore only adequate for simulated systems.

As commented in the main text, the same conclusions apply at other latencies: positive setpoints induce an undersetting of the integrator gain, while negative setpoints produce an oversetting and impose a ``gain floor'' at low S/N.
This concludes our analysis of the minimum-variance capability of CLOSE for various latencies.
We are confident this property would extend in between, to fractional latencies within that range and beyond.

\section{Experimenting with the correlation-locking setpoint -- additional data}
\label{appdx:A_r_other}

We propose in this appendix supplementary simulations to Sect.~\ref{sec:5_3_r_validation}, investigating the final long exposure SR produced on the system imager versus the setpoint $r$ used to drive CLOSE.
The experimental protocol is identical to Sect.~\ref{sec:5_3_r_validation}, generalized to three different latency values $\tau / T=0$, 1, and 2~frames, for three different AO system designs: the MICADO parameters used in the main text, as well as a PWFS and a Shack-Hartmann, both on an 8~m diameter telescope.
The parameters for those simulations, thereafter referred to as ``PWFS8M'' and ``SH8M'', are given in Table~\ref{tab:A_paramTable}, or are identical to Table~\ref{tab:5_paramTable} if unspecified.

Results with varying latencies, guide star magnitudes, seeing conditions, and wind speeds, against the $r$ setpoint, are given in Tables~\ref{tab:A_MICADO}, \ref{tab:A_PWFS8M} and \ref{tab:A_SH8M}, which also includes results presented in Sect.~\ref{sec:5_3_r_validation} (see Table~\ref{tab:A_MICADO}, ``Latency $\tau=2.T$'').
In the tables presented here, we provide for each given set of conditions [AO setup, $\tau$, $r_0$, $M_R$, wind speed $V$] the maximum SR in H-band achieved over $r$ values from -0.5 to 0.5.
This value is noted SR($r_\mathrm{max}$), and the argument at which it is attained is noted $r_\mathrm{max}$.
In the case that $r_\mathrm{max} \neq 0$, we provide the value attained for the same parameters, at our favorite all-in-one solution $r=0$; this value is noted SR$(r=0)$.

Conclusions for the MICADO simulations (Table~\ref{tab:A_MICADO}) are essentially identical to the analysis provided in Sect.~\ref{sec:5_3_r_validation}: minor fluctuations in the $r_\mathrm{max}$ value, with $r=-0.1$ or $r=0$ being almost always optimal --within the LE SR error bar of a few percents.
These simulations are therefore conclusive in showing the versatility of CLOSE for different system latencies, without additional complexity other than changing the value of $\dkc$.

The simulations performed with the PWFS8M setup (Table~\ref{tab:A_MICADO}) shows trends very similar to the MICADO setup, hinting that our conclusion that $r=0$ almost always optimally works may very well apply to a large range of PWFS designs.
A performance improvement, as well as the automation of the integrator tune-up, can also be benefited from on AO systems smaller than an ELT.
The PWFS8M simulations, at $\tau = 0$, $M_R = 12$, allow us to notice the broadening of bell function of the SR as a function of $r$, as was visible on Fig.~\ref{fig:5_rValues}.
Some $r_\mathrm{max}$ values are shifted towards positive values, up to +0.3, yet the difference between SR($r_\mathrm{Max}$) and SR($r=0$) is negligible, being less than 1\%.
This behavior is expected, given that the margin for setting $\alpha_i G_i$, thus $r$, decreases with increased latency, as the trade-off between low frequency rejection and noise amplification becomes tighter.
Thus, for very small latencies, the precision of $\alpha_i G_i$ for stable control and variance minimization is of little importance, and it follows that the setpoint $r$ also benefits from an increased margin.

The simulations with the SH8M setup, reported in Table~\ref{tab:A_SH8M}, are comparable to PWFSs, except for the brighter cases $M_R=8$.
For these, and most particularly for $M_R=8, \tau=0$, the ideal setpoint $r_\mathrm{max}$ is noticeably pushed to positive values, even hitting the extremum of our parameter space $r=+0.5$.
Although, given the recorded SRs, we may safely assume that no significant improvement would be reached for $r$ in the 0.5-1.0 range.
And yet again, the difference in performance between $r_\mathrm{max} \geq 0.5$ and $r=0$ is insignificant, except for the cases with a wind speed of 40~m.s$^{-1}$.

The cases discussed here (SH8M, $\tau=0$, $M_R=8$) are actually the one simulation where modal optimization is of least importance, if any: the sensor is linear, the S/N is extremely good, and the bandwidth effect is minimal:
the need for balance between noise amplification and turbulence rejection is quite off from the design case explored in the main text.
For all other cases with the SH8M setup, i.e., where variance minimization by the rejection/noise trade-off is valid, we validate the near-minimum variance performance of CLOSE with a null setpoint, extending the conclusions seen for PWFSs.
In particular, this also implies that the methodology proposed is sound beyond certain hypotheses of the numerical analysis of Sect.~\ref{sec:3_closeTheory}, and in particular beyond the spectrum proposed in Eq.~\ref{eq:2_3_turbSpectrum}. Indeed, the power law in Eq.~\ref{eq:2_3_turbSpectrum} ought to be changed to $-11/3$ when accounting for the aliasing with a Shack-Hartmann sensor \citep{Conan1995Wavefront}.

\begin{table}[t]
	\centering
	\caption{%
		AO numerical simulation parameters for simulation setups PWFS8M and SH8M. Unspecified parameters are identical to the MICADO setup detailed in Table~\ref{tab:5_paramTable}.
	}
	\label{tab:A_paramTable}	
	\renewcommand{\arraystretch}{1.1}
	\begin{tabular}{lll}
		\hline\hline
		\multicolumn{3}{c}{Numerical simulation configurations}\\
		& PWFS8M & SH8M\\
		\hline
		\multirow{3}{*}{Telescope} %
		& \multicolumn{2}{c}{$D$ = 8~m diameter}\\ 
		& \multicolumn{2}{c}{Circular pupil}\\
		& \multicolumn{2}{c}{(no obstruction, no spider)}\\
		\hline
		\multirow{4}{*}{DMs} %
		& \multicolumn{2}{c}{Tip-tilt mirror}\\
		& \multicolumn{2}{c}{Cartesian grid DM defined per:}\\
		& \multicolumn{2}{c}{$\quad$Pitch of 50~cm, coupling of 0.24}\\
		& \multicolumn{2}{c}{$\quad$221 controlled actuators}\\
		\hline
		WFS & \multicolumn{2}{c}{}\\
		$\quad$Wavelength & 658~nm & 500~nm\\
		$\quad$Throughput & 0.28  & 0.50\\
		$\quad$Subapertures & 24$\times$24 & 16$\times$16\tablefootmark{(a)} \\
		$\quad$Measurements & 1936 pixels total & 368 slopes total\\
		$\quad$Modulation & 4~$\frac{\lambda}{D}$ radius & - \\
		\multirow{2}{*}{$\quad$Centroiding} & - & Center of gravity\\
		&-& 300 mas/px.\\
		$\quad$Readout noise & 0.3~$e^{-}$ & \tablefootmark{(b)}3.0~$e^{-}$\\
		\hline
		RTC controller & \multicolumn{2}{c}{}\\
		$\quad$Modes controlled & 220 & 200\tablefootmark{(a)} \\
		$\quad$Frequency & \multicolumn{2}{c}{$f_S =500$~Hz}\\
		$\quad$Latency & \multicolumn{2}{c}{$\tau = [0, T, 2T]$}\\
		\hline
		\multirow{2}{*}{CLOSE} %
		& \multicolumn{2}{c}{Real-time implementation}\\
		& \multicolumn{2}{c}{$p = 0.3$; $q^\pm = 10^{-3}$; $\dkc$ = [1, 3, 5]}\\
		& \multicolumn{2}{c}{$r=0$ $[-0.5, 0.5]$}
	\end{tabular}
	\tablefoot{%
		(a) Fried geometry, with 20 highest order KL modes filtered.
		(b) The read-out noise penalty for SH is loosely based on generally available detectors given the required number of pixels. This paper has no intent of comparing Shack-Hartmann and PWFS. Guide star magnitudes for the SH simulations are adjusted for a comparable S/N.}
\end{table}

\begin{table*}
	\centering
	\caption{%
		Summary of long exposure SRs in H-band for stationary simulations exploring the latency, seeing conditions, guide star magnitude, and wind speed, against the CLOSE $r$ setpoint, for the MICADO simulation setup. $r_\mathrm{max}$: setpoint at which the maximum SR is obtained for a given parameter set; SR($r_\mathrm{max}$): maximum SR reached; SR($r=0$): SR reached for a null setpoint, in cases where $r_\mathrm{max} \neq 0$.
	}
	\label{tab:A_MICADO}
	\renewcommand{\arraystretch}{1.15}
	\begin{tabular}{ll|rrrr|rrrr|rrrr}
		\hline\hline
		\multicolumn{14}{c}{Simulation setup: MICADO}       \\
		\multicolumn{14}{c}{Latency $\tau$ = $0$}    \\
		\hline
		\multirow{2}{*}{\begin{tabular}[c]{@{}l@{}}$r_0$\\ (cm)\end{tabular}} & $M_R$                & \multicolumn{4}{c|}{12}    & \multicolumn{4}{c|}{15}    & \multicolumn{4}{c}{16}    \\
		\hline
		& V (m.s$^{-1}$)         & 10   & 20   & 30   & 40   & 10   & 20   & 30   & 40   & 10   & 20   & 30   & 40   \\
		\hline
		\multirow{3}{*}{16.3}                                                 
		& SR($r_\mathrm{max}$) & 84.0 & 83.2 & 82.7 & 81.6 & 75.5 & 69.4 & 63.8 & 58.4 & 62.4 & 46.6 & 36.3 & 30.7 \\
		& $r_\mathrm{max}$     & -0.1 & 0.0  & 0.0  & +0.1 & 0.0  & -0.1 & -0.1 & 0.0  & 0.0  & -0.1 & 0.0  & -0.1 \\
		& SR($r$=0)            & 83.9 &      &      & 81.4 &      & 68.7 & 62.8 &      &      & 45.6 &      & 28.7 \\
		\hline
		\multirow{3}{*}{12.8}                                                 
		& SR($r_\mathrm{max}$) & 74.8 & 73.3 & 72.3 & 70.3 & 61.6 & 52.0 & 44.7 & 38.3 & 43.1 & 25.3 & 17.4 & 13.8 \\
		& $r_\mathrm{max}$     & -0.1 & -0.1 & 0.0  & +0.1 & 0.0  & -0.1 & -0.1 & -0.1 & -0.1 & -0.1 & -0.1 & -0.1 \\
		& SR($r$=0)            & 74.7 & 73.2 &      & 70.2 &      & 50.0 & 42.4 & 37.5 & 41.7 & 22.8 & 15.4 & 11.8 \\
		\hline
		\multirow{3}{*}{8.9}                                                  
		& SR($r_\mathrm{max}$) & 50.4 & 48.5 & 45.0 & 41.5 & 29.6 & 20.6 & 14.3 & 10.2 & 13.6 & 4.2  & 2.2  & 0.9  \\
		& $r_\mathrm{max}$     & 0.0  & -0.1 & -0.1 & 0.0  & -0.1 & -0.1 & -0.1 & -0.1 & -0.1 & -0.1 & -0.1 & -0.1 \\
		& SR($r$=0)            &      & 47.8 & 44.6 &      & 27.8 & 16.3 & 11.2 & 6.7  & 8.3  & 1.6  & 0.8  & 0.5  \\
		\hline\hline
			\multicolumn{14}{c}{Latency $\tau$ = $T$}    \\
		\hline
		\multirow{2}{*}{\begin{tabular}[c]{@{}l@{}}$r_0$\\ (cm)\end{tabular}} & $M_R$                & \multicolumn{4}{c|}{12}    & \multicolumn{4}{c|}{15}    & \multicolumn{4}{c}{16}    \\
		\hline
		& V (m.s$^{-1}$)         & 10   & 20   & 30   & 40   & 10   & 20   & 30   & 40   & 10   & 20   & 30   & 40   \\
		\hline
		\multirow{3}{*}{16.3}                                                 
		& SR($r_\mathrm{max}$) & 82.3 & 79.4 & 74.6 & 64.4 & 72.7 & 61.6 & 50.0 & 39.4 & 58.8 & 36.6 & 23.4 & 18.1 \\
		& $r_\mathrm{max}$     &  0.0 &  0.0 & -0.1 & -0.1 & 0.0  & -0.1 & -0.1 & -0.1 & 0.0  & -0.1 & -0.1 & -0.1 \\
		& SR($r$=0)            &      &      & 74.1 & 64.1 &      & 60.4 & 47.8 & 37.8 &      & 35.9 & 21.6 & 16.9 \\
		\hline
		\multirow{3}{*}{12.8}                                                 
		& SR($r_\mathrm{max}$) & 72.0 & 66.1 & 58.7 & 47.8 & 57.7 & 41.7 & 30.1 & 21.6 & 36.9 & 19.2 & 10.2 &  6.4 \\
		& $r_\mathrm{max}$     & -0.1 &  0.0 & -0.1 & -0.1 & -0.1 & -0.1 & -0.1 & -0.1 & -0.1 & -0.1 & -0.1 & -0.1 \\
		& SR($r$=0)            & 71.8 &      & 57.7 & 44.3 & 56.4 & 39.6 & 26.9 & 20.2 & 35.9 & 13.9 &  9.2 &  5.0 \\
		\hline
		\multirow{3}{*}{8.9}                                                  
		& SR($r_\mathrm{max}$) & 46.5 & 38.0 & 29.9 & 20.5 & 24.4 & 12.7 &  6.8 &  3.8 &  9.6 &  1.9 &  0.5 &  0.3 \\
		& $r_\mathrm{max}$     & -0.2 & -0.1 & -0.1 & -0.1 & -0.1 & -0.1 & -0.1 & -0.1 & -0.1 & -0.1 & -0.1 & -0.1 \\
		& SR($r$=0)            & 45.9 & 37.4 & 26.8 & 19.0 & 21.1 & 10.2 &  4.9 &  2.2 &  7.9 &  0.8 &  0.4 &  0.2 \\
		\hline\hline
			\multicolumn{14}{c}{Latency $\tau$ = $2T$}    \\
		\hline
		\multirow{2}{*}{\begin{tabular}[c]{@{}l@{}}$r_0$\\ (cm)\end{tabular}} & $M_R$                & \multicolumn{4}{c|}{12}    & \multicolumn{4}{c|}{15}    & \multicolumn{4}{c}{16}    \\
		\hline
		& V (m.s$^{-1}$)         & 10   & 20   & 30   & 40   & 10   & 20   & 30   & 40   & 10   & 20   & 30   & 40   \\
		\hline
		\multirow{3}{*}{16.3}                                                 
		& SR($r_\mathrm{max}$) & 80.4 & 68.7 & 48.8 & 37.5 & 69.7 & 49.3 & 31.3 & 22.6 & 53.8 & 28.2 & 15.7 & 10.2 \\
		& $r_\mathrm{max}$     &  0.0 & -0.1 & -0.1 & -0.1 & -0.1 & -0.1 & -0.1 &  0.0 & 0.0  & -0.1 & -0.1 &  0.0 \\
		& SR($r$=0)            &      & 67.2 & 44.3 & 36.5 & 49.1 & 46.2 & 30.9 &      &      & 27.0 & 15.5 &      \\
		\hline
		\multirow{3}{*}{12.8}                                                 
		& SR($r_\mathrm{max}$) & 68.4 & 51.1 & 31.3 & 22.5 & 52.5 & 31.5 & 16.8 &  9.7 & 30.9 & 12.5 &  5.5 &  3.2 \\
		& $r_\mathrm{max}$     & -0.1 & -0.1 & -0.2 & -0.1 & -0.1 & -0.1 & -0.1 &  0.0 & -0.1 & -0.1 & -0.1 & -0.1 \\
		& SR($r$=0)            & 67.9 & 47.6 & 28.6 & 21.5 & 51.1 & 25.2 & 14.7 &      & 30.6 &  9.6 &  4.2 &  2.1 \\
		\hline
		\multirow{3}{*}{8.9}                                                  
		& SR($r_\mathrm{max}$) & 40.4 & 24.6 & 11.2 &  6.0 & 19.9 &  7.3 &  2.9 &  1.2 &  6.6 &  0.9 &  0.3 &  0.1 \\
		& $r_\mathrm{max}$     & -0.1 & -0.1 & -0.2 & -0.1 & -0.1 & -0.1 & -0.1 & -0.1 & -0.1 & -0.1 & -0.1 &  0.0 \\
		& SR($r$=0)            & 39.4 & 20.1 & 10.7 &  5.9 & 18.5 &  4.9 &  1.5 &  0.8 &  4.9 &  0.4 &  0.2 &      \\
	\end{tabular}
\end{table*}

\begin{table*}
	\centering
	\caption{%
		Identical to Table~\ref{tab:A_MICADO}, for the PWFS8M simulation setup, as defined in Table~\ref{tab:A_paramTable}.
	}
	\label{tab:A_PWFS8M}
	\renewcommand{\arraystretch}{1.15}
	\begin{tabular}{ll|rrrr|rrrr|rrrr}
		\hline\hline
		\multicolumn{14}{c}{Simulation setup: PWFS8M}       \\
		\multicolumn{14}{c}{Latency $\tau$ = $0$}    \\
		\hline
		\multirow{2}{*}{\begin{tabular}[c]{@{}l@{}}$r_0$\\ (cm)\end{tabular}} & $M_R$                & \multicolumn{4}{c|}{12}    & \multicolumn{4}{c|}{15}    & \multicolumn{4}{c}{16}    \\
		\hline
		& V (m.s$^{-1}$)         & 10   & 20   & 30   & 40   & 10   & 20   & 30   & 40   & 10   & 20   & 30   & 40   \\
		\hline
		\multirow{3}{*}{16.3}                                                 
		& SR($r_\mathrm{max}$) & 83.4 & 83.3 & 82.8 & 81.9 & 76.9 & 73,7 & 70.4 & 67.0 & 68.4 & 61.2 & 54.1 & 48.4 \\
		& $r_\mathrm{max}$     &  0.0 & +0.1 & +0.2 & +0.3 &  0.0 &  0.0 &  0.0 &  0.0 & 0.0  &  0.0 & 0.0  &  0.0 \\
		& SR($r$=0)            &      & 83.2 & 82.4 & 81.0 &      &      &      &      &      &      &      &      \\
		\hline
		\multirow{3}{*}{12.8}                                                 
		& SR($r_\mathrm{max}$) & 75.0 & 74.9 & 73.9 & 72.7 & 66.0 & 61.3 & 56.8 & 52.8 & 54.0 & 44.8 & 38.0 & 31.4 \\
		& $r_\mathrm{max}$     & -0.1 & +0.1 & +0.2 & +0.2 & -0.1 &  0.0 & -0.1 &  0.0 &  0.0 & -0.1 & -0.1 & -0.1 \\
		& SR($r$=0)            & 74.9 & 74.8 & 73.4 & 71.7 & 65.3 &      & 56.7 &      &      & 44.4 & 36.5 & 30.7 \\
		\hline
		\multirow{3}{*}{8.9}                                                  
		& SR($r_\mathrm{max}$) & 55.3 & 55.0 & 52.9 & 51.5 & 41.7 & 36.6 & 30.3 & 25.9 & 28.0 & 19.5 & 13.5 &  9.2 \\
		& $r_\mathrm{max}$     & -0.1 & +0.1 & +0.1 & +0.2 & -0.1 & -0.1 & -0.1 & -0.1 & -0.1 & -0.1 & -0.1 & -0.1 \\
		& SR($r$=0)            & 55.1 & 54.9 & 52.7 & 50.8 & 40.6 & 35.3 & 29.2 & 24.8 & 25.8 & 17.5 & 11.8 &  8.4 \\
		\hline\hline
		\multicolumn{14}{c}{Latency $\tau$ = $T$}    \\
		\hline
		\multirow{2}{*}{\begin{tabular}[c]{@{}l@{}}$r_0$\\ (cm)\end{tabular}} & $M_R$                & \multicolumn{4}{c|}{12}    & \multicolumn{4}{c|}{15}    & \multicolumn{4}{c}{16}    \\
		\hline
		& V (m.s$^{-1}$)         & 10   & 20   & 30   & 40   & 10   & 20   & 30   & 40   & 10   & 20   & 30   & 40   \\
		\hline
		\multirow{3}{*}{16.3}                                                 
		& SR($r_\mathrm{max}$) & 82.4 & 81.1 & 78.3 & 73.3 & 74.9 & 69.1 & 62.4 & 54.4 & 66.1 & 55.0 & 45.7 & 36.2 \\
		& $r_\mathrm{max}$     &  0.0 & +0.1 & +0.1 &  0.0 &  0.0 &  0.0 &  0.0 &  0.0 & 0.0  &  0.0 &  0.0 &  0.0 \\
		& SR($r$=0)            &      & 81.0 & 78.1 &      &      &      &      &      &      &      &      &      \\
		\hline
		\multirow{3}{*}{12.8}                                                 
		& SR($r_\mathrm{max}$) & 73.5 & 71.4 & 67.0 & 59.9 & 63.1 & 55.5 & 46.9 & 37.7 & 51.2 & 38.5 & 29.0 & 20.4 \\
		& $r_\mathrm{max}$     &  0.0 &  0.1 &  0.1 &  0.0 & -0.1 &  0.0 &  0.0 &  0.0 &  0.0 & -0.1 &  0.0 & -0.1 \\
		& SR($r$=0)            &      & 71.3 & 66.8 &      & 63.0 &      &      &      &      & 38.3 &      & 20.3 \\
		\hline
		\multirow{3}{*}{8.9}                                                  
		& SR($r_\mathrm{max}$) & 52.9 & 49.3 & 42.3 & 33.7 & 38.9 & 29.2 & 20.3 & 13.2 & 24.6 & 14.2 &  7.5 &  4.2 \\
		& $r_\mathrm{max}$     &  0.0 &  0.0 &  0.0 &  0.0 & -0.1 & -0.1 & -0.1 &  0.0 &  0.0 & -0.1 & -0.1 & -0.1 \\
		& SR($r$=0)            &      &      &      &      & 37.9 & 28.2 & 19.8 &      &      & 12.9 &  7.1 &  4.1 \\
		\hline\hline
		\multicolumn{14}{c}{Latency $\tau$ = $2T$}    \\
		\hline
		\multirow{2}{*}{\begin{tabular}[c]{@{}l@{}}$r_0$\\ (cm)\end{tabular}} & $M_R$                & \multicolumn{4}{c|}{12}    & \multicolumn{4}{c|}{15}    & \multicolumn{4}{c}{16}    \\
		\hline
		& V (m.s$^{-1}$)         & 10   & 20   & 30   & 40   & 10   & 20   & 30   & 40   & 10   & 20   & 30   & 40   \\
		\hline
		\multirow{3}{*}{16.3}                                                 
		& SR($r_\mathrm{max}$) & 81.4 & 76.6 & 65.5 & 51.2 & 72.9 & 63.3 & 50.2 & 37.6 & 63.5 & 48.4 & 35.3 & 25.3 \\
		& $r_\mathrm{max}$     &  0.0 &  0.0 &  0.0 &  0.0 &  0.0 &  0.0 &  0.0 &  0.0 &  0.0 &  0.0 &  0.0 &  0.0 \\
		& SR($r$=0)            &      &      &      &      &      &      &      &      &      &      &      &      \\
		\hline
		\multirow{3}{*}{12.8}                                                 
		& SR($r_\mathrm{max}$) & 71.7 & 64.2 & 50.1 & 34.8 & 60.7 & 47.5 & 33.6 & 21.5 & 48.0 & 31.2 & 19.6 & 12.0 \\
		& $r_\mathrm{max}$     &  0.0 &  0.0 &  0.0 &  0.0 &  0.0 &  0.0 & -0.1 &  0.0 &  0.0 &  0.0 &  0.0 &  0.0 \\
		& SR($r$=0)            &      &      &      &      &      &      & 33.2 &      &      &      &      &      \\
		\hline
		\multirow{3}{*}{8.9}                                                  
		& SR($r_\mathrm{max}$) & 49.7 & 38.9 & 24.0 & 12.7 & 35.1 & 21.5 & 10.9 &  5.3 & 20.9 &  9.1 &  4.1 &  1.9 \\
		& $r_\mathrm{max}$     &  0.0 &  0.0 & -0.1 &  0.0 & -0.1 & -0.1 & -0.1 &  0.0 & -0.1 &  0.0 &  0.0 &  0.0 \\
		& SR($r$=0)            &      &      & 23.9 &      & 34.7 & 21.4 & 10.7 &      & 20.7 &      &      &      \\
	\end{tabular}
\end{table*}

\begin{table*}[t!]
	\centering
	\caption{%
		Identical to Table~\ref{tab:A_MICADO}, for the SH8M simulation setup, as defined in Table~\ref{tab:A_paramTable}. Note: starred values ``$r_\mathrm{max} = +0.5^\ast$''indicates that the actual maximum was beyond our probing range, and thus only guarantees that $r_\mathrm{max} \geq 0.5$.
		In such cases, the value reported as SR($r_\mathrm{max}$) is the one achieved for $r=+0.5$ and not the actual maximum.
	}
	\label{tab:A_SH8M}
	\renewcommand{\arraystretch}{1.15}
	\begin{tabular}{ll|rrrr|rrrr|rrrr}
		\hline\hline
		\multicolumn{14}{c}{Simulation setup: SH8M}       \\
		\multicolumn{14}{c}{Latency $\tau$ = $0$}    \\
		\hline
		\multirow{2}{*}{\begin{tabular}[c]{@{}l@{}}$r_0$\\ (cm)\end{tabular}} & $M_R$                & \multicolumn{4}{c|}{8}    & \multicolumn{4}{c|}{11}    & \multicolumn{4}{c}{12}    \\
		\hline
		& V (m.s$^{-1}$)         & 10   & 20   & 30   & 40   & 10   & 20   & 30   & 40   & 10   & 20   & 30   & 40   \\
		\hline
		\multirow{3}{*}{16.3}                                                 
		& SR($r_\mathrm{max}$) & 80.8 & 80.8 & 81.1 & 80.9 & 76.3 & 74.0 & 72.3 & 71.1 & 63.7 & 56.3 & 50.9 & 47.7 \\
		& $r_\mathrm{max}$     & +0.3 & +0.5$^\ast$ %
										     & +0.5$^\ast$
										     		& +0.5$^\ast$
										     			   &  0.0 &  0.0 &  0.0 & +0.1 &  0.0 &  0.0 & 0.0  &  0.0 \\
		& SR($r$=0)            & 80.7 & 80.3 & 79.9 & 78.7 &      &      &      & 70.8 &      &      &      &      \\
		\hline
		\multirow{3}{*}{12.8}                                                 
		& SR($r_\mathrm{max}$) & 73.0 & 73.1 & 73.4 & 73.2 & 68.1 & 65.6 & 63.8 & 62.7 & 55.1 & 47.0 & 43.0 & 39.5 \\
		& $r_\mathrm{max}$     & +0.4 & +0.5$^\ast$
											 & +0.5$^\ast$
											 	    & +0.5$^\ast$
											 	    	   & 0.0  &  0.0 &  0.0 & +0.1 &  0.0 &  0.0 &  0.0 &  0.0 \\
		& SR($r$=0)            & 72.9 & 72.3 & 71.8 & 70.3 &      &      &      & 62.1 &      &      &      &      \\
		\hline
		\multirow{3}{*}{8.9}                                                  
		& SR($r_\mathrm{max}$) & 57.2 & 57.4 & 57.8 & 57.5 & 52.0 & 49.6 & 48.7 & 47.4 & 40.4 & 32.9 & 29.1 & 26.2 \\
		& $r_\mathrm{max}$     & +0.3 & +0.5$^\ast$
											 & +0.5$^\ast$
											 		& +0.5$^\ast$
											 			   &  0.0 &  0.0 & +0.1 & +0.1 &  0.0 &  0.0 & 0.0  &  0.0 \\
		& SR($r$=0)            & 57.0 & 56.4 & 55.5 & 53.4 &      &      & 48.1 & 46.4 &      &      &      &      \\
		\hline\hline
		\multicolumn{14}{c}{Latency $\tau$ = $T$}    \\
		\hline
		\multirow{2}{*}{\begin{tabular}[c]{@{}l@{}}$r_0$\\ (cm)\end{tabular}} & $M_R$                & \multicolumn{4}{c|}{8}    & \multicolumn{4}{c|}{11}    & \multicolumn{4}{c}{12}    \\
		\hline
		& V (m.s$^{-1}$)         & 10   & 20   & 30   & 40   & 10   & 20   & 30   & 40   & 10   & 20   & 30   & 40   \\
		\hline
		\multirow{3}{*}{16.3}                                                 
		& SR($r_\mathrm{max}$) & 80.2 & 79.1 & 77.4 & 74.0 & 74.9 & 70.4 & 66.4 & 62.8 & 61.6 & 51.9 & 44.4 & 39.3 \\
		& $r_\mathrm{max}$     & +0.3 & +0.3 & +0.2 & +0.1 & 0.0  &  0.0 &  0.0 &  0.0 & +0.1 &  0.0 &  0.0 &  0.0 \\
		& SR($r$=0)            & 80.1 & 78.6 & 76.9 & 73.9 &      &      &      &      & 61.5 &      &      &      \\
		\hline
		\multirow{3}{*}{12.8}                                                 
		& SR($r_\mathrm{max}$) & 72.2 & 70.9 & 68.7 & 64.3 & 66.2 & 61.3 & 57.3 & 52.9 & 52.4 & 42.6 & 35.8 & 30.3 \\
		& $r_\mathrm{max}$     & +0.3 & +0.3 & +0.2 & +0.1 & 0.0  &  0.0 &  0.0 &  0.0 &  0.0 &  0.0 &  0.0 &  0.0 \\
		& SR($r$=0)            & 72.0 & 70.2 & 67.9 & 64.1 &      &      &      &      &      &      &      &      \\
		\hline
		\multirow{3}{*}{8.9}                                                  
		& SR($r_\mathrm{max}$) & 56.1 & 54.3 & 51.4 & 45.6 & 50.1 & 45.1 & 40.9 & 35.9 & 37.5 & 28.2 & 22.5 & 18.2 \\
		& $r_\mathrm{max}$     & +0.3 & +0.3 & +0.2 & +0.1 & 0.0  & +0.1 & +0.1 &  0.0 &  0.0 &  0.0 &  0.0 &  0.0 \\
		& SR($r$=0)            & 55.8 & 53.4 & 50.4 & 45.3 &      & 45.0 & 40.7 &      &      &      &      &      \\
		\hline\hline
		\multicolumn{14}{c}{Latency $\tau$ = $2T$}    \\
		\hline
		\multirow{2}{*}{\begin{tabular}[c]{@{}l@{}}$r_0$\\ (cm)\end{tabular}} & $M_R$                & \multicolumn{4}{c|}{8}    & \multicolumn{4}{c|}{11}    & \multicolumn{4}{c}{12}    \\
		\hline
		& V (m.s$^{-1}$)         & 10   & 20   & 30   & 40   & 10   & 20   & 30   & 40   & 10   & 20   & 30   & 40   \\
		\hline
		\multirow{3}{*}{16.3}                                                 
		& SR($r_\mathrm{max}$) & 79.2 & 74.6 & 66.3 & 58.0 & 73.1 & 65.8 & 57.0 & 49.0 & 59.4 & 46.6 & 37.3 & 30.2 \\
		& $r_\mathrm{max}$     & +0.2 & +0.1 &  0.0 &  0.0 &  0.0 &  0.0 &  0.0 &  0.0 & 0.0  &  0.0 &  0.0 &  0.0 \\
		& SR($r$=0)            & 79.0 & 74.5 &      &      &      &      &      &      &      &      &      &      \\
		\hline
		\multirow{3}{*}{12.8}                                                 
		& SR($r_\mathrm{max}$) & 70.9 & 65.0 & 54.4 & 44.5 & 64.2 & 55.5 & 46.2 & 37.4 & 49.8 & 36.3 & 28.0 & 21.1 \\
		& $r_\mathrm{max}$     & +0.2 & +0.1 &  0.0 &  0.0 &  0.0 &  0.0 &  0.0 &  0.0 & 0.0  &  0.0 &  0.0 &  0.0 \\
		& SR($r$=0)            & 70.5 & 74.9 &      &      &      &      &      &      &      &      &      &      \\
		\hline
		\multirow{3}{*}{8.9}                                                  
		& SR($r_\mathrm{max}$) & 54.3 & 46.7 & 33.8 & 23.6 & 47.6 & 38.4 & 27.6 & 19.2 & 34.6 & 22.7 & 14.8 & 9.6 \\
		& $r_\mathrm{max}$     & +0.2 & +0.1 &  0.0 &  0.0 &  0.0 &  0.0 &  0.0 &  0.0 & 0.0  &  0.0 &  0.0 &  0.0 \\
		& SR($r$=0)            & 53.7 & 46.4 &      &      &      &      &      &      &      &      &      &     \\
	\end{tabular}
\end{table*}

\end{document}